\documentclass[twocolumn]{aastex631}

\newcommand\ACCX{\bgroup\markoverwith{\textcolor{red}{\rule[0.5ex]{4pt}{1pt}}}\ULon}

\usepackage{comment}

\begin{document}

\title{Goodbye to Chi-by-Eye: A Bayesian Analysis of Photometric Binaries in Six Open Clusters}

\author[0000-0002-9343-8612]{Anna C. Childs}
\affiliation{Center for Interdisciplinary Exploration and Research in Astrophysics (CIERA) and Department of Physics and Astronomy Northwestern University,
1800 Sherman Ave, Evanston, IL 60201 USA}

\author[0000-0002-3881-9332]{Aaron M. Geller}
\affiliation{Center for Interdisciplinary Exploration and Research in Astrophysics (CIERA) and Department of Physics and Astronomy Northwestern University,
1800 Sherman Ave, Evanston, IL 60201 USA}

\author[0000-0002-5775-2866]{Ted von Hippel}
\affiliation{Embry-Riddle Aeronautical University, Department of Physical Sciences, 1 Aerospace Blvd, Daytona Beach, FL 32114, USA}

\author[0009-0001-9841-0846]{Erin Motherway}
\affiliation{Embry-Riddle Aeronautical University, Department of Physical Sciences, 1 Aerospace Blvd, Daytona Beach, FL 32114, USA}

\author[0000-0003-3695-2655]{Claire Zwicker}
\affiliation{Illinois Institute of Technology, 10 West 35th Street Chicago, IL 60616, USA}



\begin{abstract}

We present a robust methodology for identifying photometric binaries in star clusters.  Using Gaia DR3, Pan-STARRS and 2MASS data, we self-consistently define the cluster parameters and binary demographics for the open clusters (OCs)  NGC 2168 (M35), NGC 7789, NGC 6819, NGC 2682 (M67), NGC 188, and NGC 6791.  These clusters span in age from $\sim$200 Myr (NGC 2168) to more than $\sim$8 Gyr (NGC 6791) and have all been extensively studied in the literature.  We  use the Bayesian Analysis of Stellar Evolution software suite (BASE-9) to derive the  age, distance, reddening, metallicity, binary fraction, and binary mass-ratio posterior distributions for each cluster.  We perform a careful analysis of our completeness and also compare our results to previous spectroscopic surveys. For our sample of main-sequence stars with masses between $0.6 - 1 M_\odot$, we find that these OCs have similar binary fractions that are also broadly consistent with the field multiplicity fraction.  Within the clusters, the binary fraction increases dramatically toward the cluster centers, likely a result of mass segregation. Furthermore nearly all clusters show evidence of mass segregation within the single and binary populations, respectively.    The OC binary fraction increases significantly with cluster age in our sample, possibly due to a combination of mass-segregation and cluster dissolution processes.  We also find a hint of an anti-correlation between binary fraction and cluster central density as well as total cluster mass, possibly due to an increasing frequency of higher energy close stellar encounters that inhibit long-period binary survival and/or formation.  

\end{abstract}





\section{Introduction} \label{sec:intro}

Binary and higher-order multiple stars are ubiquitous and comprise a large fraction of the stars in star forming regions \citep{ghe93,koh00,kra11,bat12,san13}, open clusters \citep[OCs,][]{mer92,pat98,pat02,gel08,gel10,hol09} and the Galactic field \citep[]{Raghavan2010,Moe2017}. The multiplicity fraction for solar-type stars in the Galactic field is approximately 50\%, and increases to $\gtrsim$70\%  for the most massive stars \citep{Raghavan2010,san12,Duchene2013,cab14}. The binary fraction\footnote{\fontsize{10}{12}\selectfont We follow the usual convention and define the binary frequency as $f_{\rm b} = N_{\rm b} / N_{\rm obj}$, where $N_{\rm b}$ is the number of binaries and $N_{\rm obj}$ is the number of ``objects" in the cluster, including both single stars and binaries.} (excluding higher-order multiples) for solar-type stars in the field is $34 \pm 2$\% \citep{Raghavan2010}. Despite their universality, work remains to understand the extent to which the birth environment affects the primordial binary frequency and global distributions of orbital parameters and mass ratios, and how these primordial characteristics are modified by dynamical processes within star clusters.  

Binaries in the Galactic field are found essentially in isolation; only the  widest field binaries (those with separations over $\sim$10$^3$ AU) are in danger of encountering passing stars, or having their orbits changed dramatically by the Galactic tidal field \citep{kai14}. However, most stars with masses $\geq$0.5~$M_{\odot}$ form in denser environments and later get dispersed into the field \citep{lad03,eva09,bre10}. Some stars are subjected to Gyrs of dynamical perturbations from their neighbors within star clusters. Though many OCs dissolve to populate the Galactic field \citep{ada01}, close stellar encounters within these birth environments can significantly modify, and even disrupt, binary systems that would otherwise be stable in the Galactic field.

Therefore, our interpretation of the observed binary populations in star clusters and the field, as well as our understanding of star formation, relies on how a population of stars and binaries evolves through this more dynamically active stage in a star cluster.  Yet we have precious little empirical data on star cluster binary properties.

Sophisticated $N$-body modeling provide predictions for the evolution of a binary population in star clusters \citep[e.g.,][]{hurley2005,por01,marks2011, geller2013b, gellernbody, geller2015b}.  Such simulations model the gravitational interactions and stellar (and binary) evolution processes for each star over Gyr timescales.  They  predict that the binary fraction in a star cluster will be modified over time by both distant two-body relaxation effects and close encounters with other single stars and stellar systems \citep[e.g.,][]{geller2013b,geller2015b}.  For instance, the effects of two-body relaxation leading to mass segregation are predicted to raise the main-sequence binary fraction in the cluster core relative to the cluster halo over time \citep{Cote1991, Layden1999, Fregeau2002}. Also, close stellar encounters are predicted to truncate the distribution of orbital separations (or periods) at the ``hard-soft" boundary \citep[e.g.][]{heggie75, gellernbody}, possibly resulting in higher-mass clusters (which generally have higher velocity dispersions) having lower binary fractions.  Stellar encounters are also expected to modify the mass-ratio distribution of a binary population \citep{marks2011, gellernbody}, for instance through exchanges leading to a mass-ratio distribution biased toward higher mass ratios. 

Observations of real OC binary populations are critical to guide such models and to verify their predictions. Traditionally, long-term spectroscopic surveys of star clusters have been employed to derive individual binary orbits, and then construct binary distributions.  Perhaps the most prolific group in this field is the WIYN Open Cluster Study \citep[WOCS,][]{mathieu2000}, providing a sample of well studied binary populations in rich open clusters of a wide range in age with a similar level of observational completeness for solar-type stars \citep[e.g.][]{gellerorbits, Geller2012, milliman_6819, leiner_m35, geller_m67, Nine2020, Tofflemire_2014}.  Astrometric, speckle, and visual binary surveys \citep[e.g.][]{Duchene2001,Patience2002,Kouwenhoven2007,Deacon2020} provide additional insights into wider binaries in OCs, while photometric variable studies \citep[e.g.][]{Mazur1995,Hartman2008,Brewer2016,Gillen2017,Brogaard2021} provide valuable information about short-period (e.g., eclipsing and ellipsoidal variable) binaries in OCs.  However these methods require multiple epochs of observations and are necessarily limited in brightness and orbital period (and also in mass-ratio, inclination and eccentricity, though to a lesser extent).
On the other hand, identifying photometric binaries requires only a single observation epoch and has no bias with respect to binary orbital period.

Unresolved binaries can be identified photometrically based on their location on a color-magnitude diagram (CMD).  Binary stars  appear redder in color and brighter than a single star with the same primary mass \citep{Hurley1998}.  Indeed, photometric binary studies have been carried out in a sample of galactic globular clusters \citep[GCs,][]{Milone2012}, and a handful of OCs \citep{binocs,Cohen2020,Jadhav2021,Cordoni2023}.  This methods requires high precision photometry and a careful isochrone fit to the cluster members.

\cite{Cordoni2023} analyzed the photometric binary populations in 78 Galactic OC's by visually determining the separation of binaries with $q>0.6$ from the MS stars in the Gaia CMD for each cluster using Gaia DR3 and previous isochrone fits from \cite{Dias2021}. \cite{Jadhav2021} use similar techniques to identify the photometric binaries in 23 Galactic OCs.  While previous studies include a much larger set of OCs for their analysis than what we consider here, we have decided to begin with a smaller sample and a more robust statistical approach using the BASE-9 code (see below). Through this paper, we develop a self-consistent analysis pipeline for identifying and characterising OCs and their photometric binary populations which we will apply to a larger sample of OCs in future papers.

As a precursor to this study, \cite{Cohen2020} used a previous version of our Bayesian analysis code, BASE-8, to analyze the photometric binaries in the OC NGC 188.  \cite{Bossini2019} used BASE-9 (an updated version of BASE-8) to determine the age, extinction, and distance for 269 OCs using Gaia DR2 photometry.  We use BASE-9 and Gaia DR3, Pan-STARRS, and 2MASS photometry to re-evaluate NGC 188 and extend our analysis to five other clusters as well.  Using complementary photometric data sets along with Gaia photometry allows us to break degeneracies in cluster parameters that exist when using only Gaia photometry \citep[mainly in distance modulus and extinction,][]{Bossini2019}.

In this paper, we describe our techniques for identifying photometric binaries in six well-studied OCs (NGC 2168, NGC 7789, NGC 6819, NGC 2682, NGC 188 and NGC 6791) that span a wide range of age and metallicity. We also use our results to compare with similar binaries in the Galactic field and globular clusters and study the effects of dynamical evolution on the binaries.  
In Section~\ref{sec:methods} we present our methods and data selection. In Section \ref{sec:BASE-9} we discuss our analysis of these data, and in Section~\ref{sec:results} we present our results.  Lastly, we provide a discussion in Section \ref{sec:discuss} and summarize our findings in Section \ref{sec:Conclusions}.



\begin{figure}[!ht]
	\includegraphics[width=.85\columnwidth]{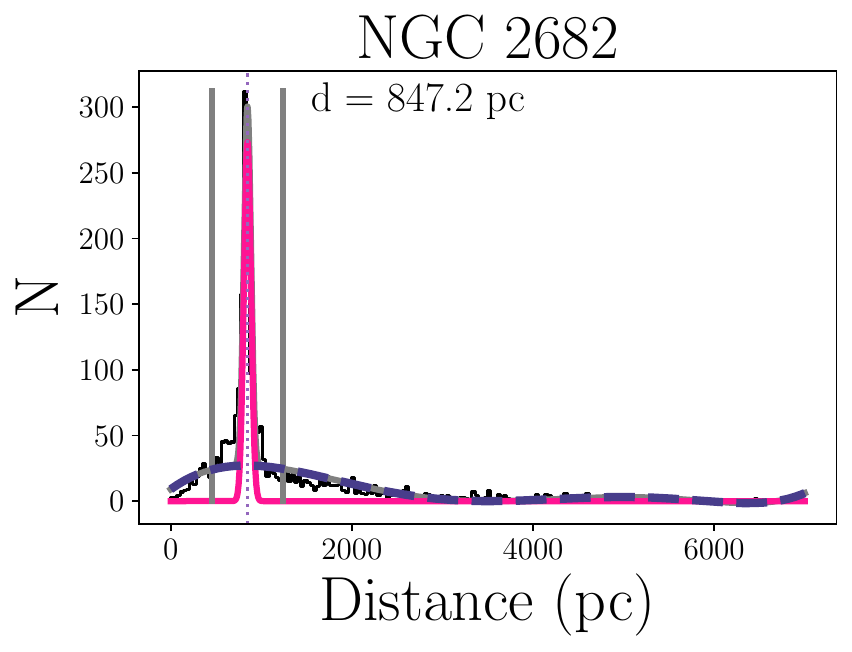}
 	\includegraphics[width=.85\columnwidth]{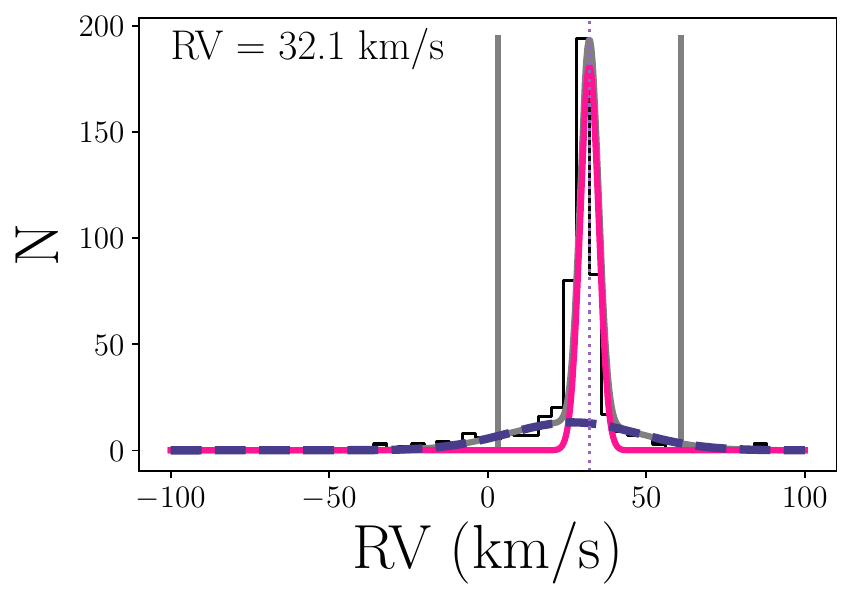}
	\includegraphics[width=\columnwidth]{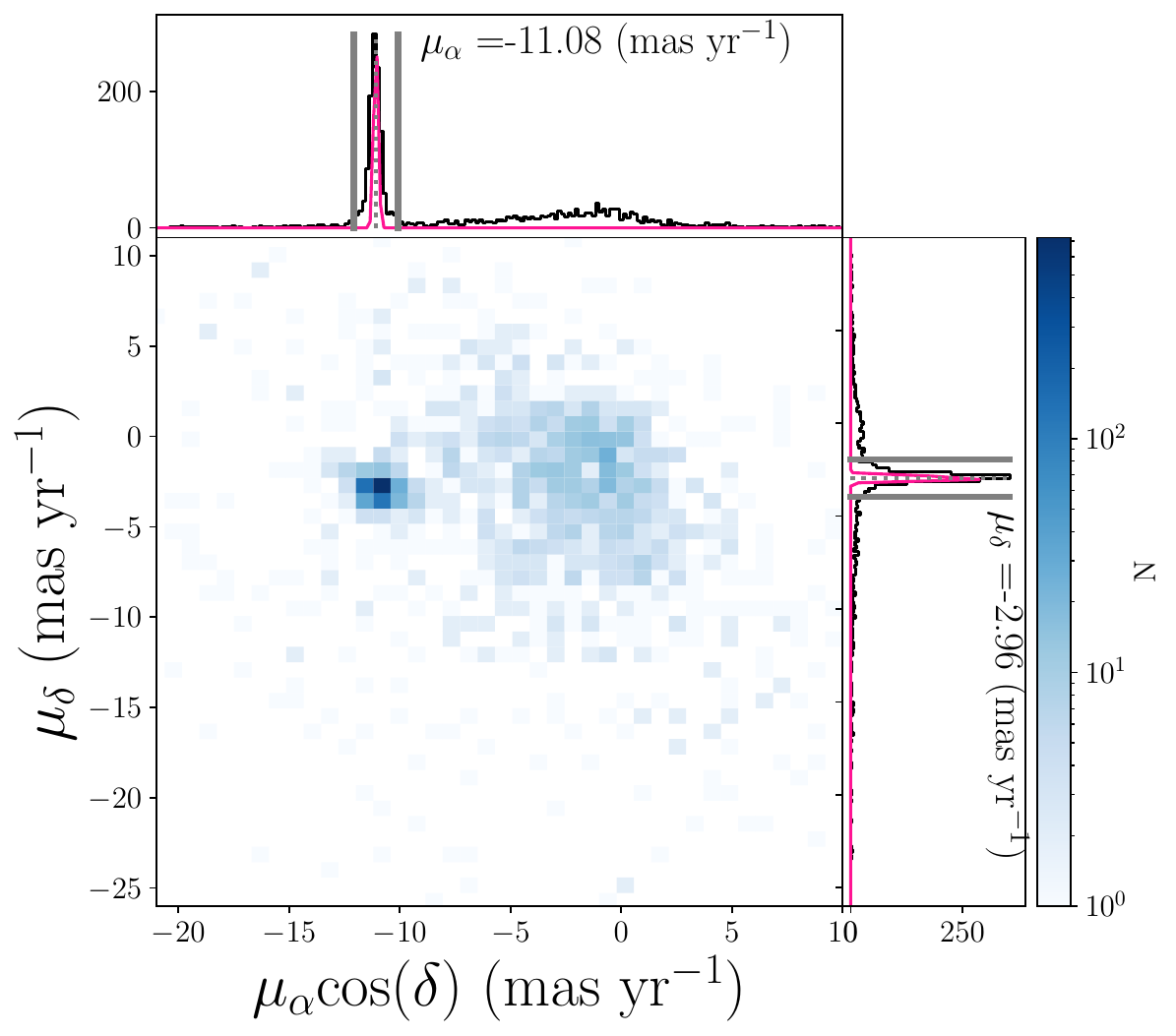}

    \caption{Cluster fits for NGC 2682 for stars within a $~0.35^{\circ}$ radius from the cluster center.  In pink we show the cluster Gaussian fit to distance (converted from parallax, top), RV (middle), and PM measurements (bottom).  In each panel, the mean of the Gaussian cluster fit is marked by a vertical dashed line, and the 10$\sigma$ bounds are marked by gray vertical lines.  A sixth order polynomial fit to the field star distance distribution and a Gaussian fit to the field RV measurements are shown by dashed purple lines, respectively.  In these top two panels, the  combined fits are shown by gray curves.} 
    \label{fig:small_fits}
\end{figure}

\section{Data and Methods} \label{sec:methods} 

In this section, we provide a summary of the data used in our study as well as our methods to recover cluster parameters and identify binaries.  For each cluster, we aim to investigate out to the cluster effective radius, $r_{\rm{eff}}$, or to 10 core radii, $r_c$ (whichever is smaller).  In short, we first use Gaia DR3 \citep{2016A&A...595A...1G, 2023A&A...674A...1G} kinematic and distance data to select a sample of probable members in each OC. For these members, we use Gaia DR3 (DOI:10.5270), Pan-STARRS \citep{chambers2019panstarrs1, Magnier_2020, Magnier_2020a, Magnier_2020b, Flewelling_2020} and 2MASS \citep{2006AJ....131.1163S} photometry (eleven different filters in total) with the Bayesian Analysis for Stellar Evolution with Nine Parameters (BASE-9) statistics software suite \citep{2006ApJ...645.1436V, 2009AnApS...3..117V, 2016ascl.soft08007R}, to characterise each cluster and recover individual stellar masses, mass ratios (for binaries) and photometric membership probabilities.  While we show our results using only Gaia photometry in Figure \ref{fig:CMDs}, BASE-9 considers all filter combinations (from each survey) when comparing to the isochrone models and determining cluster membership and stellar binarity.  We show CMDs with 2MASS and Pan-STARRS photometry in Appendix \ref{sec:CMDs}.  We use Gaia's pre-computed cross matches for both Pan-STARRS and 2MASS (see \citealt{Marrese2019} for details on cross matching the Gaia catalog with other popular catalogs).  In the following subsections, we provide further detail. cAll underlying data for this paper has been published to Zenodo (https://doi.org/10.5281/zenodo.10080762).

\subsection{Identifying cluster members from Gaia kinematics and distances}\label{ssec:membership}

We begin by downloading data from the Gaia archive out to a large radius (at least five Jacobi radii, \citep{Binney2008}) from the cluster center, which, after some initial processing, we refine to extend out to the cluster's $r_{\rm{eff}}$ (see Section~\ref{sec:results}). We calculate the Jacobi radius using cluster mass estimates from other larger OC observational surveys (see Section 2.2 of \citealt{Geller2021}).  Though the Jacobi radius is meant to approximate the cluster tidal radius, we find that the mass estimates from these surveys are typically underestimated, and therefore the resulting Jacobi radius is also underestimated\footnote{Though for the six OCs studied here there are more careful mass estimates, we use results from these larger surveys in our pipeline with an eye toward expanding to less studied clusters} (and hence we choose to multiply by a factor before downloading data to be sure to include a sufficient extent of the cluster stars).
For this sample in each cluster, we use Gaia DR3 kinematic and distance measurements as a first pass at determining which stars are likely cluster members.  For each cluster, we construct histograms of the stellar distances, radial-velocity (RV) and proper-motion (PM) measurements.  Because we are only interested in a small region on the sky, for a given cluster, we simply invert that parallax to derive distances to the stars in our sample.  Our goal is to use these cluster distributions within a pipeline that will automatically generate a reliable  membership probability for each star that will enable us to begin to separate the cluster members and field stars. Then the cluster member sample will be further refined through photometric membership analysis in BASE-9. We assume that the cluster will produce a narrow distribution, while the field will produce a broad distribution in each measurement.  Though we find this to be true for all clusters studied here, we also find that for some clusters the field can be so rich that it becomes hard to systematically distinguish the cluster peak from the field when drawing data within a large radius from the cluster center.  After some experimentation we developed a two-step process. 

First we select stars within one Jacobi radius, as defined above.  (Again, we assume these Jacobi radii are underestimated; for our purposes this conservative extent helps to limit field-star contamination). We fit functions to each distribution in distance, RV, $\mu_\alpha \cos(\rm{Dec})$ and $\mu_\delta$, respectively.  Specifically, for the distance distribution, we simultaneously fit a combined Gaussian (cluster) and polynomical (field) function. For the RV distribution, we fit a double Gaussian function (one for the cluster and one for the field).  For PM we find that we are often unable to reliably fit a combined function for the field and cluster distributions simultaneously.  We therefore only fit a Gaussian for the cluster (in each dimension, respectively) and find that this method serves our purposes.   In Figure \ref{fig:small_fits} we show an example of these fits for NGC 2682 for stars within a radius of $~0.35^{\circ}$ from the cluster center.  

Next, we apply these fits to our full sample of stars.  Specifically,  
we determine a modified $\chi^2$ value for each star based on the star's offset from the respective cluster Gaussian fit's mean value, and normalized by the standard deviation, $\sigma$, of that Gaussian,  in each dimension.  We discard any stars with any measurements (in any dimension) outside of the $10\sigma$ bound as a field star.  We arrived at a choice of $10\sigma$ through experimentation with all the clusters in our sample.  Overall, we found that choosing $3\sigma$ resulted in the exclusion of too many potential cluster members, and extending beyond $10\sigma$ did not significantly increase the sample size.  The goal at this stage is to be inclusive in our membership selection and remove the high-confidence non-members based on kinematics and distance measurements.  We then calculate the $p$-value for the $\chi^2$ statistic for each star and use this as a membership prior for BASE-9.  BASE-9 uses these priors when determining the photometric membership posteriors for each star; as decribed below, we remove photometric non-members (determined by BASE-9) before analyzing the results.

\subsection{Determining the error floor for the photometric data}


Through our initial analysis it became clear that the Gaia and Pan-STARRS photometry for the same stars are offset from each other to a level that cannot be explained by their individual uncertainties as quoted by the respective catalogs. We investigated this in two ways.  

First we used an empirical conversion formula to convert observed Gaia magnitudes for a sample of stars in our clusters to Pan-STARRS magnitudes \citep[][Ren\'{e} Andrae \textit{private communication}]{Gaia2023} in order to compare with the actual Pan-STARRS observations.  The converted Pan-STARRS magnitudes are offset from the observed Pan-STARRS magnitudes well beyond the observational uncertainties. In principle, we could use this empirical conversion between Gaia and Pan-STARRS to subtract off this shift (e.g., as a function of $g$ magnitude). However it is unclear which photometric system (Gaia or Pan-STARRS) is ``correct''. 

Second, we used PARSEC isochrones \citep[][which we use in our BASE-9 analysis]{Bressan2012}, to interpolate from Gaia to Pan-STARRS photometry for the same sample of cluster stars and compared these interpolated Pan-STARRS magnitudes to those that were observed. 
 We find that the PARSEC models  introduce an additional (unexplained) offset between the Gaia and Pan-STARRS photometry.
This is not easy to correct without modifying the models (which is well beyond the scope of this paper). 

Given these findings, we cannot know or model the photometry of any given star in BASE-9 to a precision better than 0.02 in the $g$ band (and similarly in other bands). We therefore set a floor to our uncertainty of 0.02 in all bands as an attempt to account for this offset between the photometric systems and models.

Lastly, following the methods of \cite{Motherway2023} who found an unusually large error in the $G_{\rm BP}$ band in Gaia DR3 data for a subset of stars in NGC 2168, we do not include any stars that have a $G_{\rm BP}$ error $\geq 0.13$.

\begin{table}
\begin{center}
\caption{Gaia, Pan-STARRS, and 2MASS photometric reddening coefficients for each filter.
\label{tab:reddening}
} 
\begin{tabular}{ccc}
  \hline
{Survey}  & {Filter} & {Reddening Coefficient} \\
\hline

Gaia & $G$ & 0.83627 \\
Gaia & $G_{\rm BP}$ & 1.08337 \\
Gaia & $G_{\rm RP}$ & 0.63439\\
Pan-STARRS & $g$ & 1.17994\\
Pan-STARRS & $r$ & 0.86190\\
Pan-STARRS & $i$ & 0.67648\\
Pan-STARRS & $z$ &  0.51296\\
Pan-STARRS & $y$ & 0.42905\\
2MASS & $J$ & 0.28665\\
2MASS & $H$ & 0.18082\\
2MASS & $Ks$ & 0.11675\\
\hline
\end{tabular}
\end{center}
\end{table}

\subsection{Differential Reddening}\label{ssec:red}
Differential reddening can significantly broaden the main-sequence locus, confusing efforts to identify binaries photometrically.  We attempt to correct for these effects by subtracting off the differential component of the reddening, using the three-dimensional \texttt{Bayestar19} reddening map.  \texttt{Bayestar19} includes reddening for 75\% of the sky above a declination of $-30^\circ$ \citep{Green2019}.  For a given star, we use the RA, and Dec measurements and cluster distance to determine its reddening value.  The \texttt{Bayestar19} map gives $E(B-V)$ reddening values which we then convert to $A_{\rm v}$ by the standard relation,
\begin{equation}
    A_{\rm v}=3.1 E(B-V).
\end{equation}
We adjust the Gaia, Pan-STARRS and 2MASS photometry in each passband using the $A_{\rm v}$ values from the reddening map and the reddening coefficients listed in Table~\ref{tab:reddening} \citep{Odonnell1994, Cardellin1989} to remove any differential component to the reddening.  (Then we allow BASE-9 to fit for the mean reddening value.)  We also account for the uncertainty introduced by this reddening correction by adding the error from the reddening corrections to the photometric error in quadrature. 

We find that these reddening corrections narrow the main-sequence locus and lead to better results from our BASE-9 analysis for most of the clusters in our sample.  However, the uncertainties associated with the reddening for NGC 6819 and NGC 7789 introduce photometric uncertainties that are too large for BASE-9 to derive precise cluster parameters, and so we do not apply reddening corrections to these two clusters.  We suspect this may slightly inflate the binary fractions in these two clusters, though likely not beyond the uncertainties we derive on their binary fractions.  See Section \ref{sec:diff_red} in the Appendix for more discussion on this.


\begin{figure}[!t]
	\includegraphics[width=1\columnwidth]{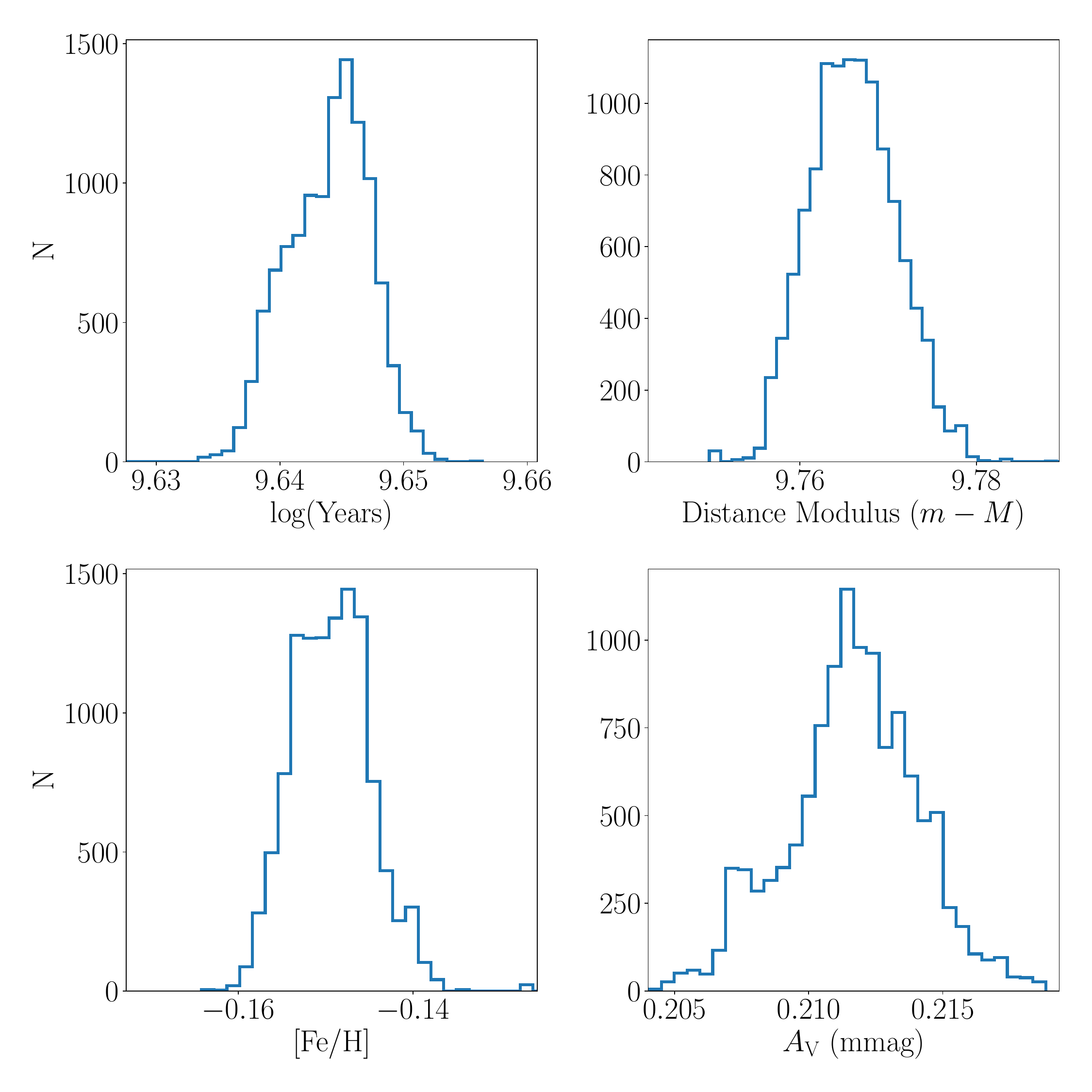}
    \caption{Posterior distributions for NGC 2682 parameters.} 
    \label{fig:posteriors}
\end{figure}

\begin{figure*}[!ht]
	\includegraphics[width=2\columnwidth]{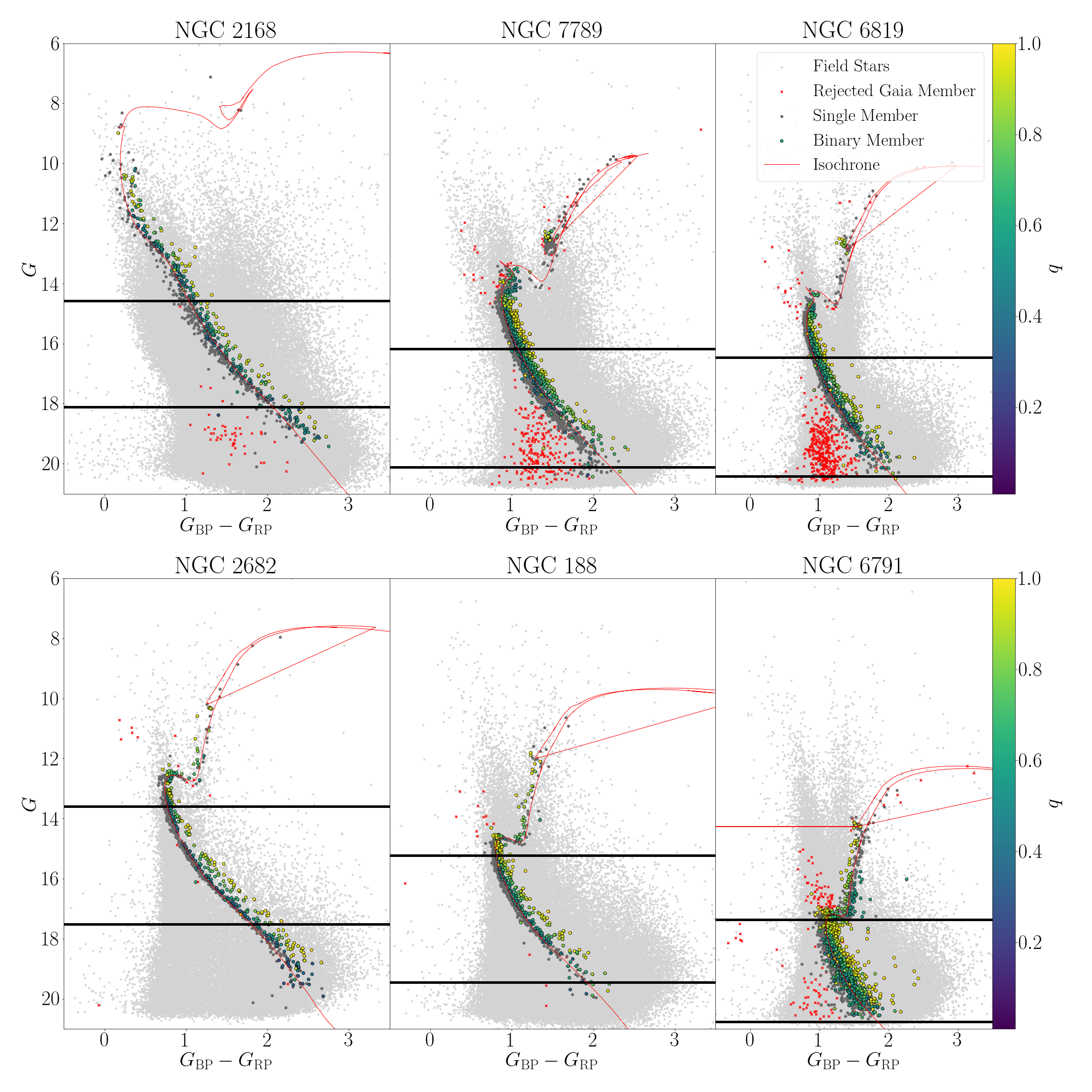}
    \caption{CMDs with Gaia photometry showing BASE-9 results for each cluster.  Field stars are marked in light gray and stars classified as members using Gaia kinematics and distances but are rejected by BASE-9 are marked by red 'x's.  Single star BASE-9 members are marked in dark gray and BASE-9 binary members are colored according to the mass ratio of the binary.  The black horizontal bars mark the magnitude range in which MS stars with a primary mass of $0.6 \, M_{\odot} \leq M_1 \leq 1 \, M_{\odot}$ are found.  The red line shows a PARSEC isochrone created from the median cluster parameters from our BASE-9 analysis.} 
    \label{fig:CMDs}
\end{figure*}

\section{BASE-9 analysis}\label{sec:BASE-9}
BASE-9 is a Bayesian software suite used to analyze photometric observations in relation to stellar evolution models and returns cluster-wide posterior distributions in absorption (Av), age, distance, and metallicity ([Fe/H]), and star-by-star posterior distributions in (primary) mass, mass ratio (for binaries) and photometric membership.  In short, BASE-9 uses priors for the global cluster parameters, stellar masses and cluster membership values, then employs a Markov Chain Monte Carlo (MCMC) method to sample parameter space and return these posterior distributions \citep{2006ApJ...645.1436V, 2009AnApS...3..117V, 2016ascl.soft08007R}.

We include the priors that we used for each cluster parameter in Table \ref{tab:Priors}.  The prior sigma on age is flat in log(Age) over the age range of the models.  Our method for determining the stellar membership priors is provided in Section~\ref{ssec:membership}.  BASE-9 uses a \citet{Salpeter1955} Initial Mass Function as the prior for stellar mass, and the user provides a starting mass value for each star.  We find that the results are  much less sensitive to the starting value of stellar mass (than, e.g., the priors on cluster parameters and stellar membership), and therefore we simply set each star's mass starting value to 1$M_\odot$.  For more information on how these priors are used within BASE-9, please see the references above.   
We choose to use the PARSEC stellar evolution models \citep{Bressan2012} for this analysis, and provide BASE-9 with the eleven filters of photometric data (and their uncertainties) described above.

For our purposes here, BASE-9 operates in two stages.  First it derives the global cluster parameters, using the \texttt{singlePopMcmc} function.  At this step, BASE-9  integrates out the star-by-star parameters, providing a significant speed-up.  We provide an example of the posterior distributions derived from BASE-9 for this step for NGC 2682 in Figure \ref{fig:posteriors}.  In most cases, the distributions are roughly symmetric and Gaussian. We also visually inspect the sampling histories for each parameter in each cluster to ensure that BASE-9 is reliably covering parameter space.  For summary statistics, we choose to provide the median values and the 16th and 84th percentiles (the equivalent to 1$\sigma$ below and above the median) of the posterior distribution for each parameter.  We use these median values for the global cluster parameters to generate the PARSEC isochrones shown in Figure~\ref{fig:CMDs}.  The median value and the $16^{\rm th}$ and $84^{\rm th}$ confidence intervals of the posterior distributions for these parameters are also listed in the first four rows of Table \ref{tab:systems}.  

Note that the uncertainties provided in Table~\ref{tab:systems} only show the $1\sigma$ width of the respective posterior distributions.  They do not account for uncertainties intrinsic to the stellar evolution models, or those introduced by the choice of photometric filters used here \citep[see][]{Hills2015}, or any other source of uncertainty.  Therefore, these uncertainties likely do not fully describe the true precision with which we know each parameter.  As one particular example, many detailed (spectroscopic) studies have found NGC 2682 to have a near solar metallicity \citep[e.g.,][]{Tautvaisiene2000,Randich2006, Pancino2010}.  However, the metallicity derived for NGC 2682 with BASE-9 is significantly sub-solar (perhaps because of underestimated uncertainties).  Nonetheless, the isochrone appears to describe the data well enough for our purposes (Figure~\ref{fig:CMDs}).   

Next, BASE-9 uses the entire posterior distribution of each global cluster parameter to evaluate the star-by-star masses, mass-ratios, and photometric  membership posterior distributions with the \texttt{sampleMass} function.  This recovers the posterior distributions of the parameters that were integrated out during the \texttt{singlePopMcmc} step while still utilizing each step in the \texttt{singlePopMcmc} sampling history that define the cluster parameters.  Again, we use the median and the $16^{\rm th}$ and $84^{\rm th}$ confidence intervals as summary statistics of the posterior distributions to determine membership, binarity, and masses for each star.  Specifically, we require a median photometric membership value $\geq$0.01 to consider a star a member. We also follow \citet{Cohen2020} and consider a given star a binary if the median value of the posterior distribution in secondary mass is $\geq3\sigma$ from zero. 

In Figure~\ref{fig:CMDs} we show the results from this BASE-9 analysis in the Gaia filters, though note that BASE-9 uses all combinations of all eleven filters to determine the posterior distributions.  We show CMDs for other filter combinations in Appendix \ref{sec:CMDs}.

Finally, we note that the 2MASS data for these clusters typically have significantly larger errors than the Gaia and Pan-STARRS data for the same stars.  Within BASE-9, a dataset with large errors (like the fainter stars in 2MASS data) won't be as constraining on model parameters as a dataset with smaller errors.  Therefore the brighter stars in 2MASS data will provide more useful constraints on the cluster parameters than the fainter 2MASS stars.  To better understand how 2MASS data affects our results we reran our pipeline for the six clusters omitting 2MASS photometry.  We find that the posterior distributions of the cluster parameters (e.g., Figure~\ref{fig:posteriors}) are similar, indicating that the sampling is largely determined by the more precise and numerous Gaia and Pan-STARRS photometry.  When determining cluster membership and stellar binarity using photometry without 2MASS, BASE-9 finds $\geq$98\% of the members from our standard analysis pipeline in each cluster but finds only 75\%-95\% the number of binaries.  This is because 2MASS extends the SEDs and can place tighter constraints on stellar binarity for the brighter stars (with smaller 2MASS photometric uncertainties), especially for low-$q$ binaries which may otherwise be classified as single stars by BASE-9.

\subsection{Testing for Incompleteness}\label{sec:incompleteness}

Binaries with low mass ratio, $q$, occupy a similar area on the CMD as single stars.  Therefore we assume that BASE-9 will be incomplete for binaries with low $q$.  Indeed \citet{Cohen2020} found that the BASE-8 results for NGC 188 become incomplete for $q \lesssim 0.5$, using a very similar method to our work.  In order to test and correct for incompleteness in our results, we generate synthetic photometric data sets for each cluster where the single stars, binary stars, and their masses are well defined.  We then run the synthetic data set through BASE-9 to test the binary recovery.

We follow a similar method to \cite{Cohen2020} to generate synthetic cluster data. Specifically, we use the \texttt{simCluster} function in BASE-9 to generate main-sequence stars along each of the median cluster isochrones with 50\% binary fraction drawn from a uniform mass-ratio distribution sampled between $(0.08,1.00)$.  For each cluster we generate enough synthetic stars to match the number of main-sequence stars in the observational data within the defined mass range.

After generating our synthetic main-sequence stars, we add synthetic noise to the photometry in each filter for each of the three surveys.  To do this, we bin the observational photometric errors in bins of 0.1 magnitude in each filter, and we calculate the mean and standard deviation for each bin.  We then offset the synthetic photometry in each filter by a random draw from a normal distribution centered at zero and with a standard deviation equal to that of the photometry bin for the corresponding error. An example of the synthetic noise derived from this method can be seen in Figure 3 of \cite{Cohen2020}.  An example of the synthetic data for NGC 188 is shown in CMDs in Appendix \ref{sec:CMDs}.

\begin{figure}
	\includegraphics[width=1\columnwidth]{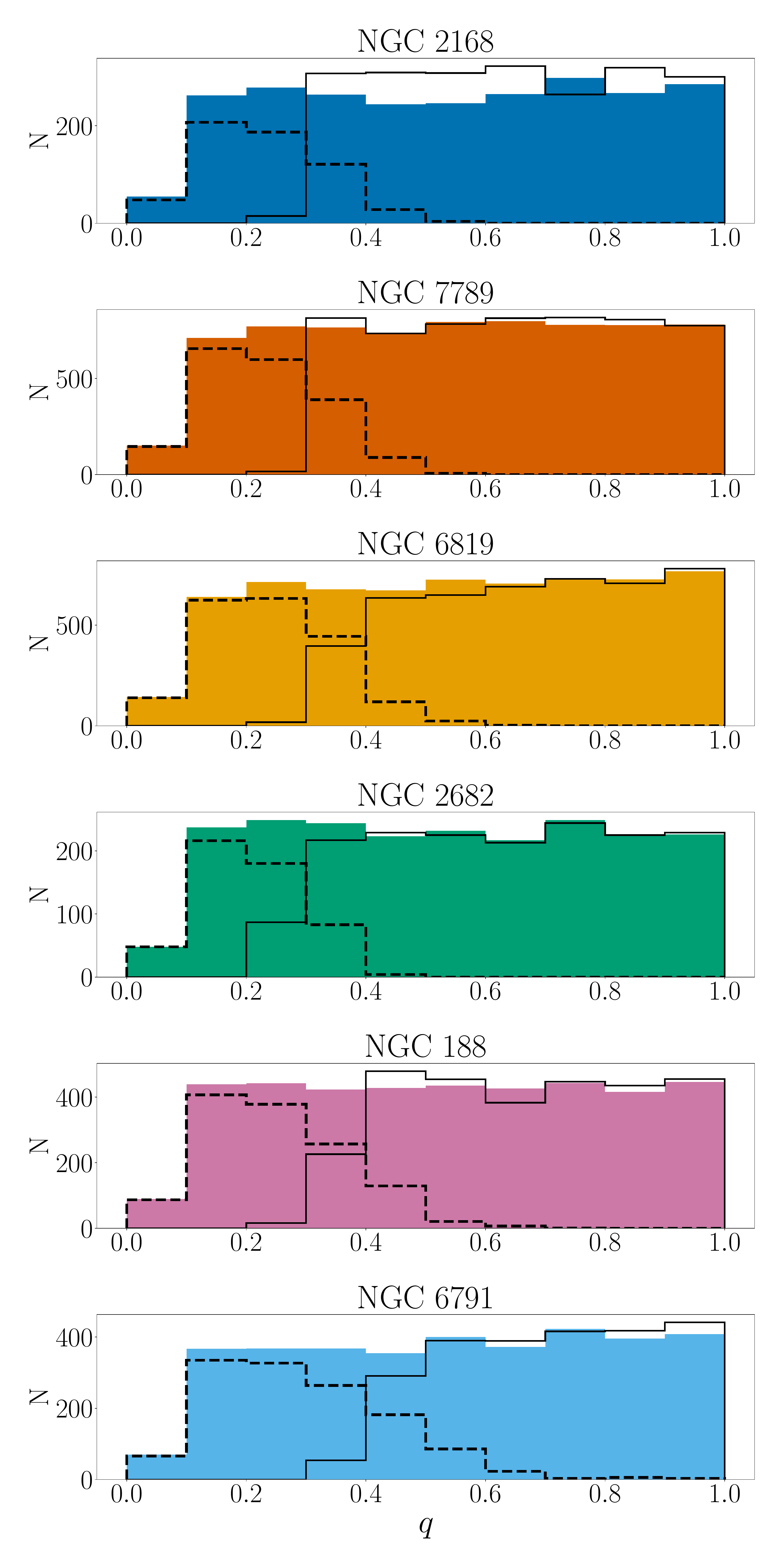}
    \caption{Simulated (color), recovered (black lines), and missed (dashed lines) binaries for the ML-MS stars with $0.6 \, M_{\odot} \leq M_1 \leq 1 \, M_{\odot}$ in bins of mass ratio.} 
    \label{fig:sim_q}
\end{figure}

For each cluster, we run ten sets of synthetic photometric data (e.g., 10 synthetic clusters) through \texttt{sampleMass} and using the \texttt{singlePopMcmc} posterior distributions we derived with the real data for that cluster to evaluate BASE-9's binary recovery rate.  The clusters each sample a slightly different range in $M_1$, and therefore we will restrict our discussion here to $0.6 \, M_{\odot} \leq M_1 \leq 1 \, M_{\odot}$ (e.g., see Figure~\ref{fig:CMDs}) so as to compare like samples across clusters.  We report the results in Table \ref{tab:simulation_results}.  The column $f\textit{(bin,in)}$ reports the known binary fraction of the synthetic data set.  The column $f\textit{(bin,out)}$ reports the recovered binary fraction from BASE-9 and $f\textit{(rec)}$ is the binary recovery rate.  
We also perform these same statistical tests but only for binaries with $q\geq 0.5$.  Similar to \citet{Cohen2020}, we find our recovery rate of binaries with $q \geq 0.5$ is nearly 100\% in all clusters.  

We also explore the incompleteness as a function of $q$.  Figure \ref{fig:sim_q} shows a histogram of the simulated binary $q$ for each cluster.  Black solid lines show the $q$ distributions of the binaries recovered by BASE-9, and black dashed lines show the $q$ distributions of the binaries that were missed by BASE-9 (and identified as single stars).  We see that ($i$) BASE-9 consistently identifies low-$q$ binaries as single stars and ($ii$) for $q \gtrsim 0.4$ BASE-9 is relatively complete.  
Though not shown in this figure, we also find that ($iii$) BASE-9 is more consistent in recovering the primary mass than $q$, and ($iv$)  for high $q_{\rm in}$ BASE-9 often returns a lower $q_{\rm out}$, and the inverse is true for low $q_{\rm in}$. We return to this in Section~\ref{ssec:qdists}.

We use these simulation results and recovery rates to correct our statistics in our subsequent analysis.



\begin{table*}
\begin{center}
\caption{Cluster priors and locations.
\label{tab:Priors}
} 
\begin{tabular}{c|cccc|cc}
  \hline
{Cluster}  & {log(Age)/yrs} & {m-M} & {[Fe/H]} &{$A_{\rm V}$/mmag} & RA & Dec \\

\hline

NGC 2168 & $8.098 $ & {$9.644 \pm 0.103$} & {$-0.16 \pm 0.30$} & {$0.834 \pm 0.300 $} & $6^{\rm h}09^{\rm m} 07.5^{\rm s}$ & $+24^{\circ}20'28"$  \\

NGC 7789 & $9.204 $ & {$11.290 \pm 0.300$} & {$0.02 \pm 0.50$} & {$0.702 \pm 0.500 $} & $23^{\rm h}57^{\rm m} 21.6^{\rm s}$ & $+56^{\circ}43'22"$  \\

NGC 6819 & $9.285 $ & {$12.120 \pm 0.175$} & {$0.05 \pm 0.30$} & {$0.760 \pm 0.300 $} & $19^{\rm h}41^{\rm m} 17.5^{\rm s}$ & $+40^{\circ}11'47"$ \\

NGC 2682 & $9.470$ & {$9.640 \pm 0.101$} & {$-0.046 \pm 0.30$} & {$0.180 \pm 0.300 $}  & $8^{\rm h}51^{\rm m} 23.3^{\rm s}$ & $+11^{\circ}49'02"$  \\

NGC 188 & $9.608 $ & {$11.284 \pm 0.123$} & {$-0.030 \pm 0.30$} & {$0.267 \pm 0.300 $}  & $0^{\rm h}47^{\rm m}$ & $+85^{\circ}15'$  \\

NGC 6791 & $9.718 $ & {$12.964 \pm 0.015$} & {$0.23 \pm 0.30$} & {$0.374 \pm 0.300 $}  & $19^{\rm h}20^{\rm m} 53^{\rm s}$ & $+37^{\circ}46'30"$  \\

\hline
\end{tabular}
\end{center}
\end{table*}

\begin{table*}
\begin{center}
\caption{Cluster parameters.
\label{tab:systems}
} 
\begin{tabular}{c|cccccc}
  \hline
{}  & {NGC 2168} & {NGC 7789} & {NGC 6819} & {NGC 2682}  & {NGC 188}  & {NGC 6791} \\
\hline
{Age/Gyr} & 0.213$^{+0.0001}_{-0.0002}$  &  1.778$^{+0.0015}_{-0.0013}$ &  2.428$^{+0.0040}_{-0.0045}$ & 4.407$^{+0.0310}_{-0.0408}$ &  6.167$^{+0.0124}_{-0.0185}$ & 8.444$^{+0.0058}_{-0.0032}$ \\

{Distance/kpc} & 0.826$^{+0.0025}_{-0.0022}$ &  1.853$^{+0.0008}_{-0.0007}$&2.419 $^{+0.0008}_{-0.0005}$ & 0.814$^{+0.0016}_{-0.0012}$ & 1.816$^{+0.0014}_{-0.0010}$ &  4.168$^{+0.0008}_{-0.0014}$\\

{[Fe/H]} &-0.058$^{+0.0066}_{-0.0057}$ & 0.025$^{+0.0029}_{-0.0029}$ & -0.035$^{+0.0026}_{-0.0029}$ & -0.150 $^{+0.0048}_{-0.0044}$ &  0.018$^{+0.0060}_{-0.0046}$ &  0.235$^{+0.0016}_{-0.0021}$\\

{$A_{\rm V}$/mmag} &713.75$^{+3.990}_{-4.270}$ & 852.04$^{+2.006}_{-2.139}$ & 556.47$^{+2.019}_{-2.148}$ & 211.77$^{+2.292}_{-2.166}$ &144.10$^{+2.755}_{-2.939}$ &   521.96$^{+1.7470}_{-0.8890}$\\

{$R_{\rm h}^{\circ}$} & $0.42$ & $0.43$ &  $0.17$ &$0.25$ & $0.17$ & $0.10$\\
$R_{\rm h}$/pc & 5.98 & 13.68 &  7.10 & 3.57 &5.34 &  7.34\\
{$r_{\rm eff}$} &  60.0$^{\prime}$ & 30.0$^{\prime}$ &25.0$^{\prime}$ &  65.0$^{\prime}$ &35.0$^{\prime}$ & 20.0$^{\prime}$\\
{$r_{\rm c}$} &   13.49$\pm 1.30 ^{\prime}$ & 13.87$\pm 0.84 ^{\prime}$ & 5.47$\pm 0.43^{\prime}$ & 8.17$\pm 0.69 ^{\prime}$ & 5.48$\pm 0.28^{\prime}$ & 3.28$\pm 0.08^{\prime}$\\
{$r_{\rm t}$} &  77.46$\pm 4.21^{\prime}$ & 36.05$\pm 0.65^{\prime}$ & 33.37$\pm 1.05^{\prime}$ & 84.57 $\pm 5.46^{\prime}$ & 40.97 $\pm 1.36^{\prime}$ &  23.49 $\pm 0.31^{\prime}$\\ 
$M/M_{\odot}$ &  {$1622_{-62} ^{+56}$} &  {$4954_{-130} ^{+105}$}& {$3136 _{-85} ^{+64}$}& {$1269_{-35} ^{+29}$}& {$1282_{-44} ^{+32}$}& {$4939_{-211} ^{+144}$}\\
$N$ & {$1232$} & {$4011$} & {$2632$} &{$1283$} & {$1140$} &  {$4141$}\\
$N_{\rm c}$ & {$507$} & {$2642$} &  {$1139$} & {$298$} &  {$388$} &  {$1403$}\\
{$t_{\rm rh}$/Myr} & $295^{+65}_{-40}$ & $1525^{+215}_{-139}$ &  $506^{+83}_{-54}$ & $159^{+30}_{-17}$ & $263^{+38}_{-15}$ &  $617^{+50}_{-22}$ \\
Age/$t_{\rm rh}$ & 0.7  & 1.3 & 4.8 & 27.7 & 23.4 & 13.7  \\

\hline
\end{tabular}
\end{center}
\end{table*}

\begin{table*}
\begin{center}
\caption{Simulation results.
\label{tab:simulation_results}
} 
\begin{tabular}{ccccccc}
  \hline
{Cluster}  & {\textit{f(bin,in)}} & {\textit{f(bin,out)}} & {\textit{f(rec)}} & {\textit{f(bin,in,q$\geq$0.5)}} & {\textit{f(bin,out,q$\geq$0.5)}} & {\textit{f(rec,q$\geq$0.5)}} \\
\hline
NGC 2168 & $0.505 \pm 0.003$ & $0.454 \pm 0.005$ & $0.900 \pm 0.010$ & $0.279 \pm 0.005 $ & $0.323 \pm 0.007$ & $1.157 \pm 0.017$ \\

NGC 7789 & $0.510 \pm 0.005$ & $0.402 \pm 0.005$ & $0.789 \pm 0.009$ & $0.284 \pm 0.004 $ & $0.289 \pm 0.004$ & $1.018  \pm 0.004$ \\

NGC 6819 & $0.512 \pm 0.005$ & $0.360 \pm 0.003$ & $0.704 \pm 0.005$ & $0.288 \pm 0.004 $ & $0.278 \pm 0.004$ & $0.967 \pm 0.003$ \\

NGC 2682 & $0.500 \pm 0.007$ & $0.388 \pm 0.008$ & $0.776 \pm 0.011$ & $0.267 \pm 0.006 $ & $0.264 \pm 0.007$ & $0.986 \pm 0.007$ \\

NGC 188 & $0.500 \pm 0.005$ & $0.355 \pm 0.006$ & $0.708 \pm 0.007$ & $0.272 \pm 0.005 $ & $0.265 \pm 0.005$ & $0.976 \pm 0.009$ \\

NGC 6791 & $0.517 \pm 0.005$ & $0.348 \pm 0.007$ & $0.673 \pm 0.011$ & $0.293 \pm 0.006 $ & $0.297 \pm 0.008$ & $1.015 \pm 0.009$ \\

\hline
\end{tabular}
\end{center}
\end{table*}

\begin{figure*}
	\includegraphics[width=2\columnwidth]{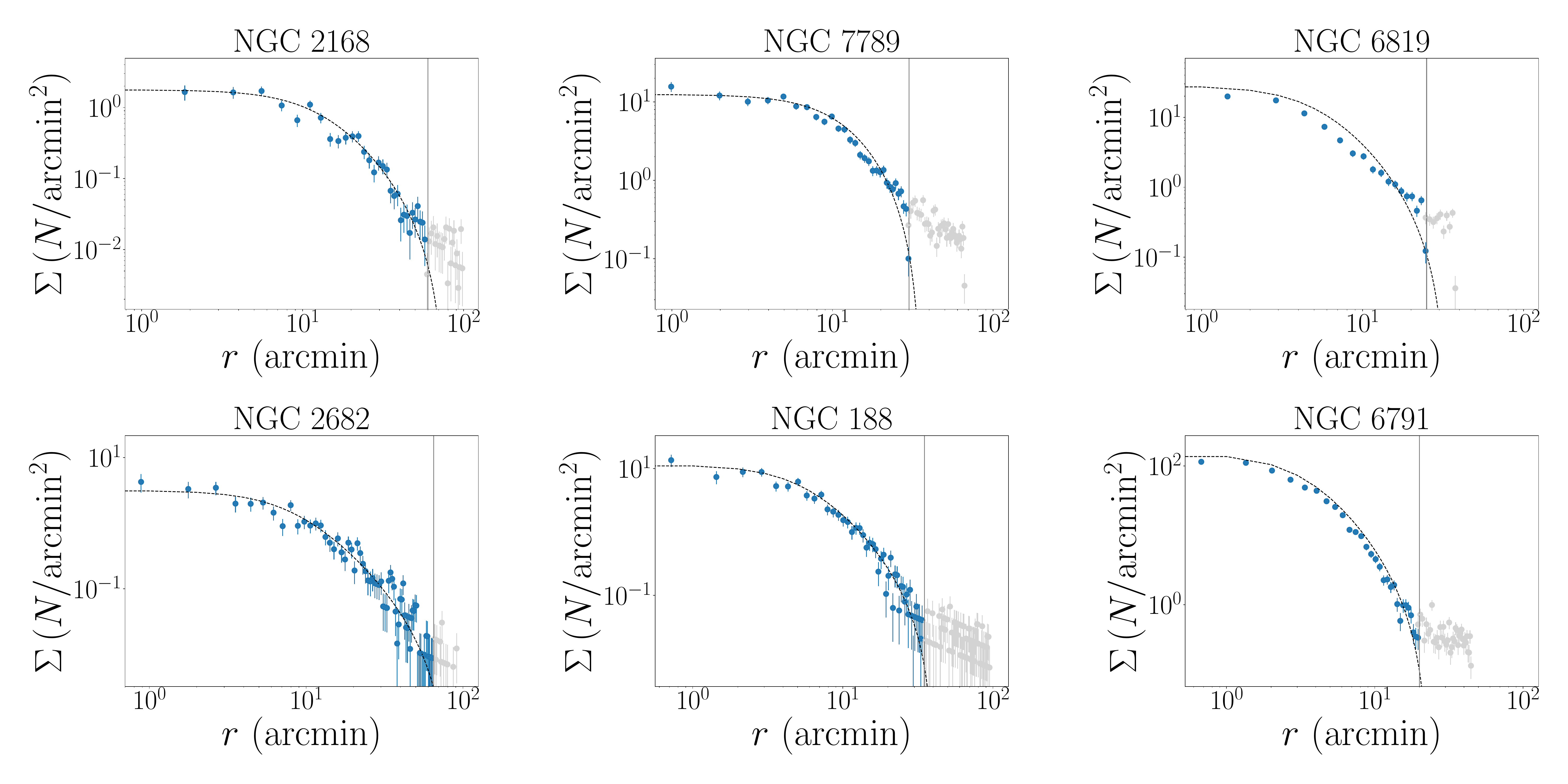}
    \caption{\citet{King1962} model fits using only ML-MS stars within $r_{\rm eff}$. 
 We plot the observed (binned) surface density profile in points with error bars, and we show the fit in the black dotted line.   We mark the effective radius by a gray vertical line in each panel, and plot the stars that we do not consider in our analysis (found beyond $r_{\rm eff}$) in gray.} 
    \label{fig:Kingfits}
\end{figure*}

\section{Results}\label{sec:results}

In the following section we present results that we derive from our analysis of this sample of OCs.  The interested reader can access and download these data and analysis products, and create their own custom interactive figures and tables using these data on our ``Open Cluster Binary Explorer'' website: \url{http://ocbinaryexplorer.ciera.northwestern.edu/}. 

\subsection{Cluster CMDs}

Figure \ref{fig:CMDs} shows the BASE-9 results for each of our six clusters. Note again that though we show the data here in the Gaia filters, BASE-9 uses all combinations of the eleven filters we provided to determine the cluster and star-by-star parameters.  Our pipeline achieves a self-consistent fit across all clusters.  We see in general that the median isochrones describe the observations well for all clusters, and as expected $q$ generally increases toward the red side of the isochrone.  

However, we also notice some important discrepancies between the observations and the isochrones.  First, the giant branch is typically not fit well by these models.  In particular, the base of the giant branch in the isochrones do not always reach the apparent base of the giant branch in the observations.  Second, BASE-9 appears unable to reliably determine mass ratios of binaries near the turnoff for clusters that have a ``blue hook'' morphology in the isochrone, e.g., for NGC 7789 and NGC 6819.  It is not clear if this is a limitation from BASE-9 or from the stellar evolution models themselves.  Further exploration of this is beyond the scope of this paper.  (Also, the blue stragglers and other binary evolution products in all clusters that are known to lie far from a standard isochrone are rejected by BASE-9 as field stars, though this is not surprising.)  Nonetheless, the majority of the main-sequence stars appear to be well described by this BASE-9 analysis. 

In order to make a direct comparison between clusters, we restrict our analysis to main-sequence stars with a primary mass $0.6 \, M_{\odot} \leq M_1 \leq 1 \, M_{\odot}$, unless stated otherwise.  This is the largest mass range of MS stars available in the photometric data for the oldest cluster we consider, NGC 6791.  We refer to this subset of cluster stars as the mass-limited - main-sequence stars (ML-MS).  We mark the magnitude range that corresponds to these mass limits with black horizontal bars for each cluster.

\subsection{King model fits and effective radii}

Figure \ref{fig:Kingfits} shows the stellar surface density versus distance from cluster center for each OC.  Here we include only stars that are within the ML-MS sample.  Beyond some effective radius, $r_{\rm eff}$, the surface density flattens.  This is likely due to a low cluster surface density toward the outer edges of the cluster and contamination from field stars.  To remove this contamination we limit all our analysis to stars within $r_{\rm eff}$.  We estimate the $r_{\rm eff}$ value by eye and include the value for each OC in Table~\ref{tab:systems}.

To constrain the core radius ($r_{\rm c}$) and the tidal radius ($r_{\rm t}$) of each cluster we fit to stellar surface density ($\Sigma$) a \citet{King1962} model of the form 
\begin{equation}
    f(x)=A r_{\rm c}^2 \left ( \frac{1}{\sqrt{1 + (r/r_{\rm c})^2}} - \frac{1}{\sqrt{1 + (r_{\rm t}/r_{\rm c})^2}}   \right )^2,
\end{equation}
where $A$ is a constant and $r$ is the distance from the cluster center.  We only include the ML-MS stars within $r_{\rm eff}$ for this analysis.  The resulting fits are shown in Figure~\ref{fig:Kingfits}, and the corresponding $r_{\rm c}$ and $r_{\rm t}$ for each cluster are listed in Table \ref{tab:systems}.

We note that the surface density profile and the resulting values for $r_{\rm c}$ and $r_{\rm t}$ depend on the stellar sample used.  Therefore, it is difficult to compare directly with values from the literature that use different sample selection criteria. In general, we find larger core radii in most clusters than previous studies \citep[e.g.][]{Mathieu1983, Wu2009, Kang2002, Chumak2010} , except for NGC 188 and NGC 6791 \citep[][where we agree]{Bonatto2005, Tofflemire_2014}.  
We suspect these differences are due to variations in the stellar samples between our MS-ML mass range and the various selection criteria in the literature.

\subsection{Cluster timescales}\label{ssec:trelax}
The half-mass relaxation time of the cluster can be approximated by
\begin{equation}
    t_{\rm rh}=\frac{0.346 N (r_{\rm c,3D})^{3/2}}{\sqrt{GM}ln \Lambda},
\end{equation}
where $N$ is the total number of objects (single or higher order star systems) in the cluster, $G$ is the gravitational constant, $M$ is the total mass of the cluster and $\Lambda$ is Coulombs constant which we assume to be $0.1 N$ \citep{Giersz1994,Heggie2003}.  $r_{\rm c,3D}$ is the deprojected core radius, equal to $4/3 r_{\rm c}$ \citep{Spitzer1987}.  Note that this equation is typically written in terms of the half-mass radius, $R_{\rm h}$.  However, as we do not sample the entire cluster radial profile (due to low cluster surface density at large radii), we are unable to reliably derive $R_{\rm h}$ directly from our data.  Instead we approximate the half-mass radius with $R_{\rm h}=1.846 r_{\rm c,3D}$ \citep{Heggie2003} and report that value in Table~\ref{tab:systems} in both degrees and parsecs.

We estimate the cluster values for $N$ and $M$  using all likely cluster members as determined by BASE-9, regardless of primary mass, and list these values in Table~\ref{tab:systems}.  We suspect that both of these numbers are underestimated as our analysis does not reach the tidal radius of any cluster, and our photometry does not reach to the faintest cluster members.   In Table \ref{tab:systems} we also report $N_{\rm c}$, the number of objects found within one core radius from the cluster center.

\begin{table*}
\begin{center}
\caption{Binary fraction of the ML-MS stars in each cluster.  The uncorrected, raw binary fractions from BASE-9 are listed in the $f_{\rm b, i}$ columns and the corrected BASE-9 binary fractions are listed in the $f_{\rm b, c}$.  The binary fractions from the WOCS spectroscopic survey, $f_{\rm W}$, the fraction of the WOCS binaries that were recovered, $f_{\rm W,r}$, the fractions of the WOCS double lined binaries that we recovered $f_{\rm W,DL}$, and the reference from which the RV data are from are also listed.
\label{tab:binary_results}
} 
\begin{tabular}{c|cc|ccc|c}
  \hline
{Cluster}  &{$f_{\rm b,i}$} & {$f_{\rm b,c}$} & {$f_{\rm W}$} & {$f_{\rm W,r}$} & {$f_{\rm W,DL}$}  &  {Ref.} \\
 & (BASE-9) &(BASE-9) & (WOCS) &  & &  \\
\hline
NGC 2168 & 0.38 $\pm$ 0.03 & 0.43 $\pm$ 0.03  & 0.24 $\pm$ 0.03 & 33/38 & 5/5 &  \cite{Leiner2015}\\

NGC 7789 & 0.23 $\pm$ 0.01 & 0.29 $\pm$ 0.01 & 0.31 $\pm$ 0.04 & 20/85 & 3/15 & \cite{Nine2020} \\

NGC 6819 &  0.31 $\pm$ 0.01&  0.44 $\pm$ 0.02 & 0.22 $\pm$ 0.03 & 24/72 & 9/14  & \cite{Miliman2014} \\

NGC 2682  & 0.36 $\pm$ 0.03 & 0.46 $\pm$ 0.03 & 0.34 $\pm$ 0.03 & 62/97 & 23/29 & \cite{Geller2021}\\

NGC 188 & 0.38 $\pm$ 0.02 & 0.52 $\pm$ 0.03  & 0.29 $\pm$ 0.03 & 50/107 & 10/12  &  \cite{Geller2009} \\ 

NGC 6791 & 0.41 $\pm$ 0.01 & 0.60 $\pm$ 0.02  & - & - & - & \cite{Tofflemire_2014}  \\

\hline
\end{tabular}
\end{center}
\end{table*}

Using these values we calculate $t_{\rm rh}$ and the number of dynamical timescales each cluster has undergone, $\rm{Age}/t_{\rm rh}$.  This is likely an underestimate of the true $t_{\rm rh}$ for each cluster (due to underestimating $N$ and $M$), but it does provide a guide to understanding how dynamically relaxed the cluster is expected to be.  The younger clusters in our sample have only experienced $\sim$1-5 relaxation times, while the older clusters have experienced 10s of relaxation times.

\begin{figure*}[!t]
	\includegraphics[width=2\columnwidth]{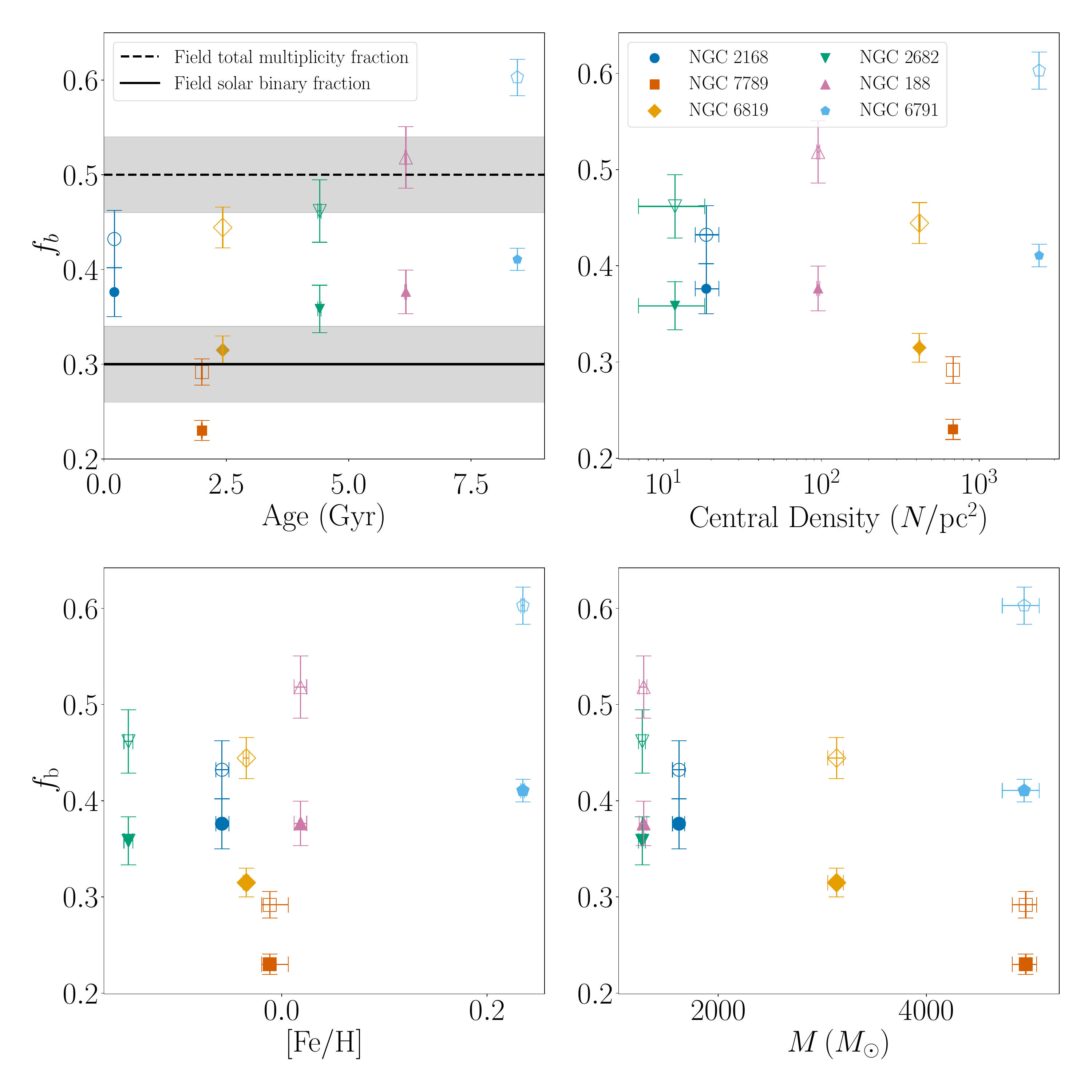}

    \caption{ML-MS binary fraction vs. the cluster age (upper-left), central density (upper-right), [Fe/H] (lower-left) and total cluster mass (lower-right). The closed circles are the uncorrected binary fractions ($f_{\rm b,i}$) and the open circles are the corrected binary fractions ($f_{\rm b,c}$).  The colors correspond to the cluster in the same manner as previous figures.  In the upper-left panel, the black line denotes the field binary fraction for solar type stars, the black dashed line denotes the field total multiplicity fraction, and the gray region marks the uncertainty of these binary fractions.} 
    \label{fig:fb}
\end{figure*}

\subsection{Comparing binary fractions against cluster-wide parameters}\label{ssec:bexternal}
In Table \ref{tab:binary_results} we report the cluster-wide binary fractions from our analysis and the literature for each cluster in our sample.  Specifically, we include the raw (incomplete) binary fractions ($f_{\rm b,i}$) of the ML-MS sample calculated directly from the BASE-9 output, the corrected binary fractions of the ML-MS ($f_{\rm b,c}$) after adjusting $f_{\rm b,i}$ using the recovery fraction $f(rec)$ as determined by the simulations in Table \ref{tab:simulation_results}, the MS spectroscopic binary fractions from WOCS ($f_{\rm W}$), the fraction of WOCS spectroscopic binaries we recovered ($f_{\rm W,r}$), the fraction of WOCS doubled-lined spectroscopic binaries we recover ($f_{\rm W,DL}$), and the references for the WOCS binary data.  (There are no WOCS data for MS binaries in NGC 6791 as these stars are too dim.)  Note that the uncertainties quoted for our binary fractions are simple Poisson counting errors and the denominators of $f_{\rm W,r}$ and $f_{\rm W,DL}$ are only the WOCS stars that we identified as likely cluster members from Gaia kinematics and submitted to BASE-9.

Given that the WOCS RV survey is sensitive to lower binary mass ratios (and our BASE-9 method is sensitive to longer orbital periods), we do not expect to recover all the WOCS binaries. The WOCS binaries that BASE-9 did not recover are either ($i$) rejected as Gaia non-members (e.g., for WOCS clusters that did not have proper-motion data available at the time of publication), ($ii$) rejected as BASE-9 non-members because of their CMD location (e.g., the blue stragglers) or ($iii$) have mass ratios low enough that BASE-9 did not detect the secondary star (and therefore we labelled them as singles).  We find the biggest discrepancy between the clusters for which we were unable to correct for differential reddening, NGC 7789 and NGC 6819.  This suggests that our methods may be highly sensitive to extinction although further tests are needed to verify this.  

The WOCS spectroscopic survey is only sensitive to binaries with orbital periods less than $10^4 \, \rm days$ \citep{Geller2012}, while we expect that much wider binaries exist in each cluster.  
Except for NGC 7789, we find that $f_{\rm b, i}$ is higher than the WOCS spectroscopic binary fractions for all clusters, and correcting for incompleteness we find $f_{\rm b, c}$ to be much larger (more than $1\sigma$) than the WOCS binary fractions.  This result suggests that there is a substantial number of binaries with periods beyond $10^4 \, \rm days$ in these clusters.  We discuss NGC 7789 in more detail in Section~\ref{sec:discuss}.

In Figure \ref{fig:fb}, we plot the binary fraction against various cluster-wide parameters. In all panels, we show $f_{\rm b, i}$ in solid circles and $f_{\rm b, c}$ in open circles, with the point color defined by the cluster as in previous figures.  First we look at the top-left panel for a comparison to the field binary fraction.  We mark the solar-type field binary fraction, $0.3\pm 0.04$, by a black line and gray band and total multiplicity fraction (including all higher ordered systems), $0.5\pm 0.04$, by a black dashed line and gray band (see Table 13 of \citealt{Moe2017}).  This field binary fraction is for all binaries with $0.8 \, M_{\odot} \leq M_1 \leq 1.2 \, M_{\odot}$ and $q>0.1$, and has been corrected for various observational biases.  We see that nearly all of the raw binary fractions for our clusters are either consistent with or higher than the field value. Furthermore, nearly all of the incompleteness corrected binary fractions from our clusters are above the field binary fraction and more closely consistent with the field total multiplicity fraction.  As BASE-9 makes no distinction about higher-order systems (and only works from the combined light of a given system) it is possible that some of the stars that we label as binaries are actually members of higher-order multiples (though with low enough mass ratios that they still reside within the binary locus on the CMD). 

Turning to trends with cluster parameters, we find in our (limited) sample of OCs that $f_{\rm b}$ increases with cluster age but is generally uncorrelated with cluster central density, metallicity, and total mass.  Specifically, fitting a linear function to the incompleteness corrected binary fraction vs.\ cluster age (and accounting for the errors on the binary fractions), we find a slope of $0.04 \pm 0.01$, significantly above zero, and a Pearson correlation coefficient of 0.79.  We also tested for a trend in $f_{\rm b}$ with age excluding NGC 7789 and NGC 6791 (which have the lowest and highest $f_{\rm b}$ values, respectively), and find a slope that is still $>3\sigma$ above zero.  

A similar analysis of the other panels in Figure~\ref{fig:fb},including all OCs in our sample, return slopes that are consistent with zero and Pearson correlation coefficients near zero.  However, we note that NGC 6791 is the oldest, most metal rich, densest, and nearly the most massive (second to NGC 7789) cluster in our sample and therefore exerts a strong weight on these correlations.  If we exclude NGC 6791 from the analysis, we find that $f_{\rm b}$ is significantly anti-correlated with central density and cluster mass

We provide some speculative discussion and interpretation of these results in Section~\ref{sec:discuss}, but note that our sample size is small.  The focus of this paper is to test our methods and analysis so that these same techniques may be scaled up to a much larger sample size in future work, where we will have stronger statistical grounds to interpret any remaining  correlations in the context of cluster formation and evolution.




\begin{figure*}[!t]
	\includegraphics[width=2\columnwidth]{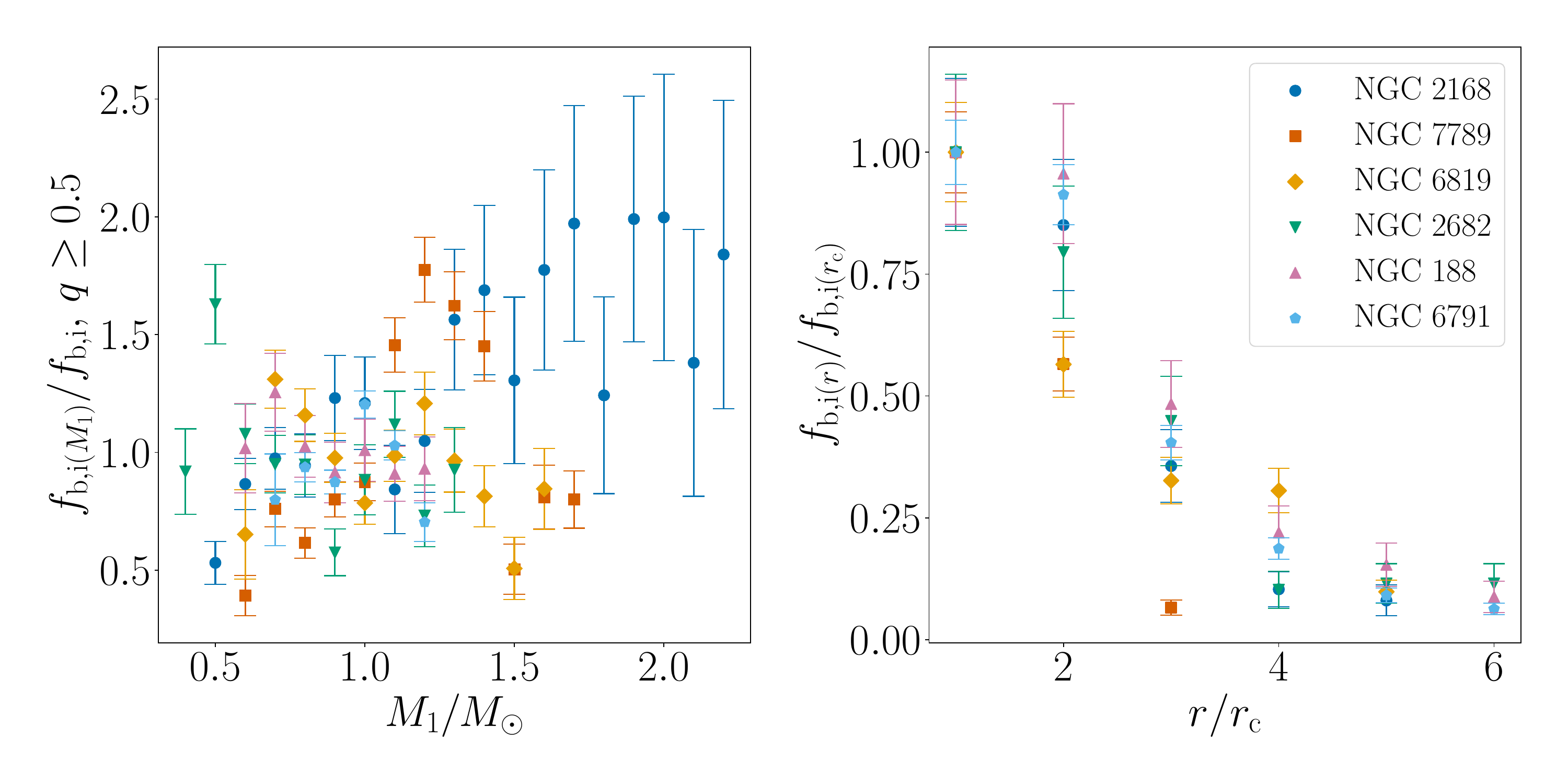}

    \caption{Binary fraction plotted against primary-star mass (left) and radius from the cluster center (right).  On the left, we show the raw binary fraction for a given $M_1$ bin for binaries with $q\geq 0.5$, $f_{\rm b, i}(M_1)$, normalized by the total raw binary fraction of the cluster for binaries with $q\geq 0.5$, $f_{\rm b,i}$.  On the right we plot binary fraction in bins of core radii away from the cluster center, $f_{\rm b,i}(r)$, normalized by the binary fraction within one core radius, $f_{\rm b,i}(r_1)$. } 
    \label{fig:all_fbs}
\end{figure*}

\subsection{Comparing binary fractions against internal cluster parameters}\label{ssec:binternal}

In the left panel of Figure \ref{fig:all_fbs} we plot the raw (incomplete) binary fraction as a function of $M_1$ for all clusters (no longer limited to be within the ML-MS sample).  The binary fraction for each $M_1$ bin is normalized by the total raw binary fraction, $f_{\rm b,i}$, for binaries with $q \geq 0.5$. We show $1\sigma$ error bars for each data point. Visually there appears to be a trend of increasing binary fraction with increasing primary mass. However, statistically if we fit a line to these data, we find a slope of $0.14 \pm 0.11$, consistent with zero, and a Pearson's correlation coefficient of 0.36.  Interestingly, the slope we find for our OC data is similar to that of \citet{Milone2012} who find a slope of $0.196 \pm 0.282$ for their photometric binaries in GCs.  In the field, there is a pronounced increase in multiplicity fraction with primary-star mass
\citep{Raghavan2010, Duchene2013}, though over a much larger mass range.  We return to this in Section~\ref{sec:discuss}.

In the right panel of Figure \ref{fig:all_fbs} we show the raw binary fraction as a function of distance from the cluster center for all stars, $f_{\rm b, i}(r)$, normalized by the binary fraction within one core radius,  $f_{\rm b, i}(r_{\rm c})$, for each cluster.  The binary fraction clearly increases toward the respective cluster centers.  A similar result was also reported by \cite{Milone2012} in their sample of GCs.

As an additional check on the radial distribution of the binaries, we plot in Figure \ref{fig:MLMS_CDFs} the normalized cumulative distribution functions (CDFs) of the singles (black) and binaries (red) as functions of distance from cluster center in units of core radius for ML-MS sample. For each cluster, we perform a Kolmogorov-Smirnov (K-S) test between the single star and binary star distributions and provide the resulting $p$-values in the figure.  In all clusters but NGC 188 and NGC 7789, we find significantly low $p$-values, suggesting that the single and binary samples are drawn from different parent distributions, respectively.  For these clusters, the binaries are more centrally concentrated than the single stars.  Interestingly, for NGC 188 \citet{gel08} also found only a marginal distinction between the radial distributions of their RV sample of solar-type binary and single stars, despite the old dynamical age of the cluster.  

Figure \ref{fig:M1_CDFs} has a similar format to Figure~\ref{fig:MLMS_CDFs} but with the singles and binaries both separated into two different mass samples.  The solid lines denote singles (red) or binaries (black) with $M_1 \geq 0.8 \, M_{\odot}$ and the dashed lines denotes singles or binaries with $M_1 < 0.8 \, M_{\odot}$.  We perform K-S tests between each $M_1$ population and report the $p$-values for both in the figure.
We find that the two mass bins for the binaries are significantly distinct in four of the six clusters, with the higher-mass binaries being more centrally concentrated in all clusters except NGC 188 and NGC 7789.  For the single stars, the more massive sample is significantly centrally concentrated in all clusters except NGC 2168 and NGC 6791.  (These findings confirm results also presented in \cite{Motherway2023} and Zwicker et al., 2023 \textit{submitted} using the same data.)

Overall, we find that for most clusters there is clear evidence that the binaries occupy a more centrally concentrated spatial distribution than the single stars and that within the single and binary populations the more massive objects are more centrally concentrated.  We return to this in Section~\ref{sec:discuss}.

\subsection{Binary mass-ratio distributions} \label{ssec:qdists}

\begin{figure}
\includegraphics[width=1\columnwidth]{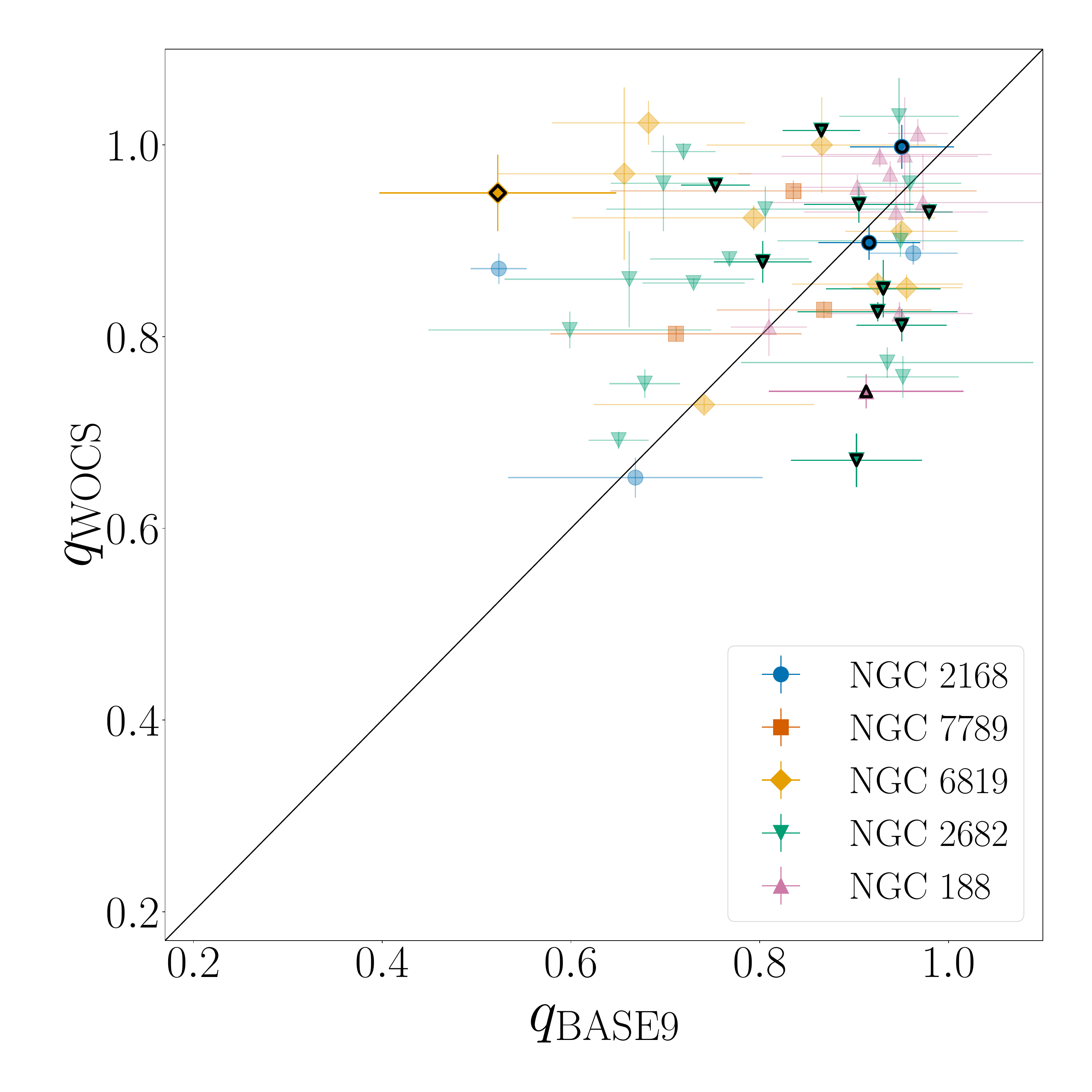}
    \caption{$q$ from the double lined binaries in WOCS vs $q$ recovered by BASE9 with $1\sigma$ error bars for all clusters.  The black diagonal line is a 1:1 line for reference. Dark points show binaries within the ML-MS sample, and lighter points show those outside the sample.  } 
    \label{fig:DL_binaries}
\end{figure}

In Figure~\ref{fig:DL_binaries} we compare our recovered photometric $q$ values to the spectroscopic $q$ values for binaries that have double-lined RV measurements from the WOCS survey.  We see that for these high-$q$ spectroscopic binaries, BASE-9 can underestimates $q$, especially in younger clusters.   The mean difference between the WOCS and BASE-9 $q$ values for the double-lined binaries in our ML-MS sample is $\sim$0.13.  This is larger than the predicted difference in recovered $q$ from our simulations using only BASE-9.  The reason for this discrepancy between spectroscopic and BASE-9 photometric $q$'s is not entirely clear but was also reported in \citet{Cohen2020}.  As pointed out in \cite{Cohen2020}, the posterior distribution for $q$ is (by definition) truncated at 1, and therefore for high $q$, the median (and mean) value will likely not be at the peak of the distribution which will result in a lower $q$ value of the summary statistic (e.g., the median, as we report here).  For the purposes of this paper, we assume that we cannot know the $q$ value of any individual binary to better than 0.2. 

\begin{figure}
	\includegraphics[width=0.95\columnwidth]{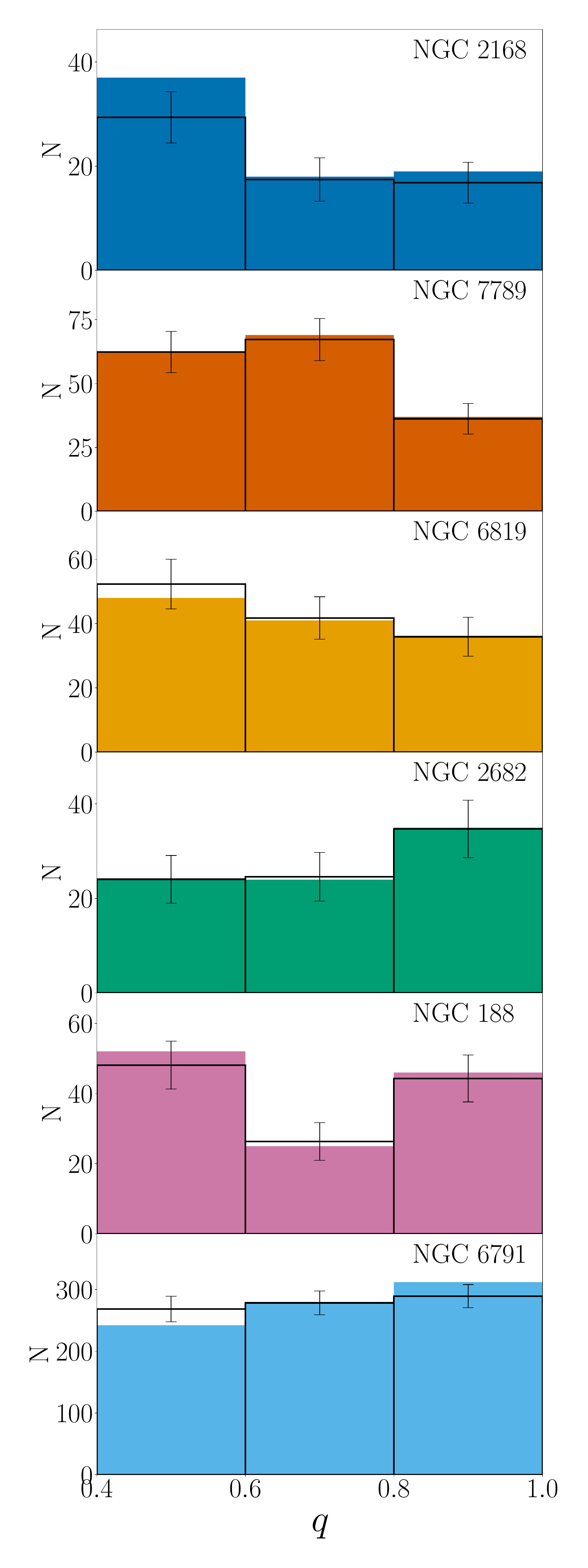}

    \caption{Histograms of $q$ for ML-MS binaries with $q > 0.4$. The corrected values and the uncertainties for each $q$ bin are marked with black lines. } 
    \label{fig:q_hist}
\end{figure}

\begin{figure*}
	\includegraphics[width=2\columnwidth]{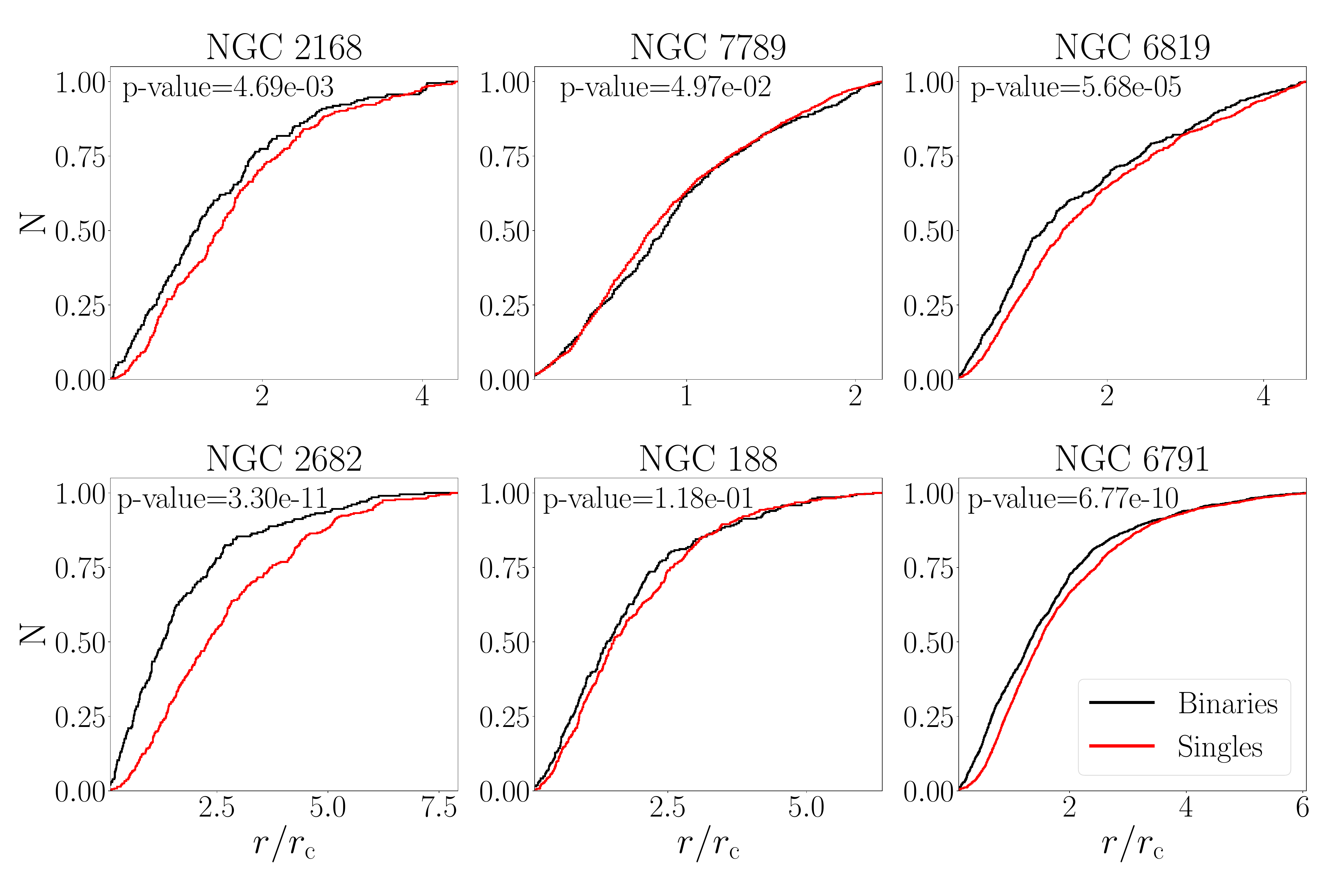}
    \caption{Normalized CDFs of all ML-MS binaries in black and ML-MS single stars in red, vs distance from cluster center in units of core radii.} 
    \label{fig:MLMS_CDFs}
\end{figure*}

\begin{figure*}
	\includegraphics[width=2\columnwidth]{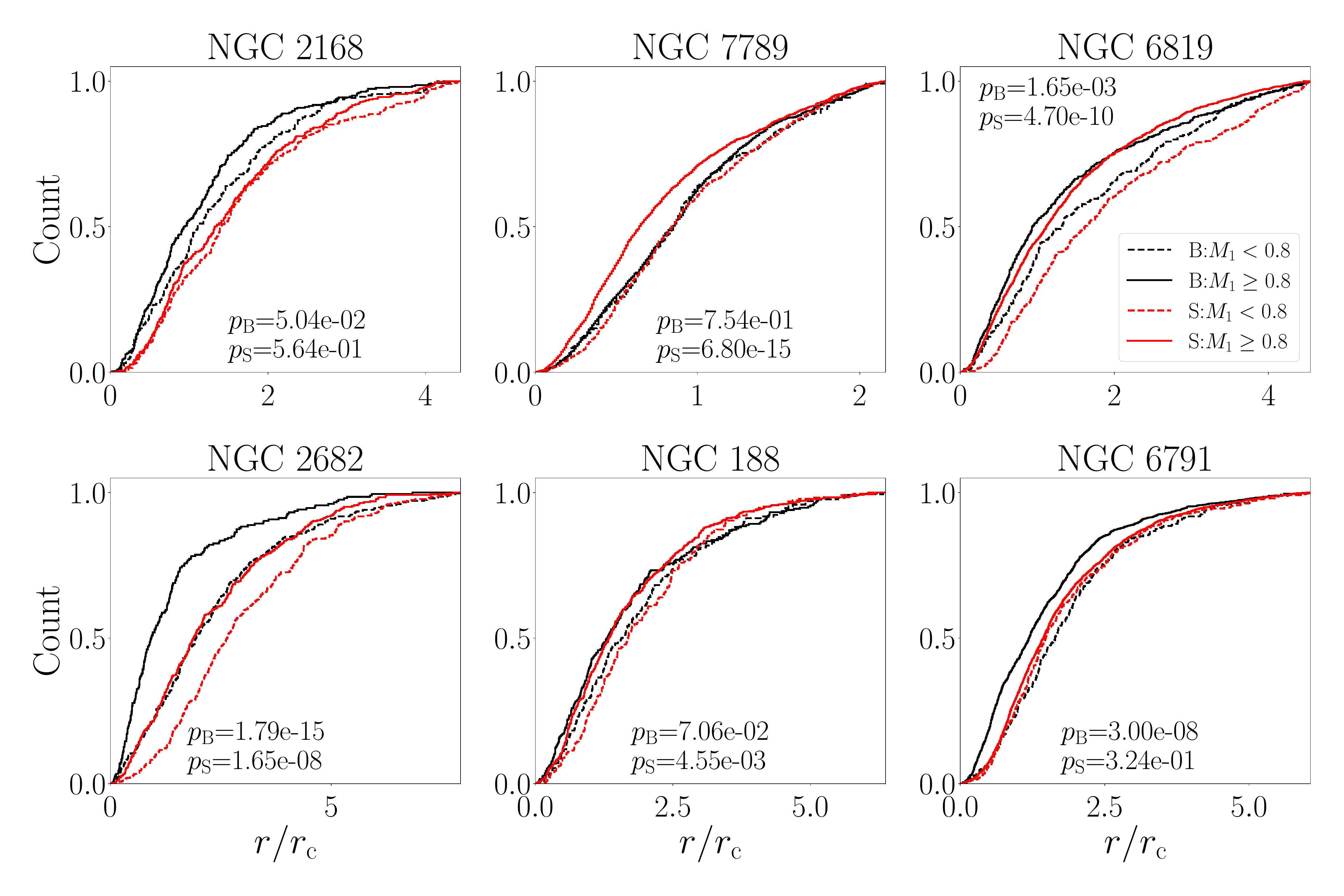}
    \caption{Normalized CDFs of the radial distributions of our photometric binaries (black) and single stars (red) in units of core radii. We divide the binaries and single stars both into two groups: the solid lines denote objects with $M_1 \geq 0.8 \, M_{\odot}$ and the dashed lines denote objects with $M_1 < 0.8 \, M_{\odot}$.} 
    \label{fig:M1_CDFs}
\end{figure*}

In Figure \ref{fig:q_hist} we show histograms of the ML-MS binaries with $q > 0.4$ (within bins of width 0.2). The incompleteness corrected values and the uncertainties for each $q$ bin are marked with black lines. All clusters have $q$ distributions within this sample that are consistent with being uniform.  This agrees with RV results for NGC 2682 and NGC 188 \citep{Geller2012, Geller2021}, the only two clusters where $q$ distributions from spectroscopic binaries were analyzed in the literature.  (Though the mass ranges in those studies are not identical to ours, they do substantially overlap with ours, and therefore we would not expect a significant difference in $q$ distributions between our sample and theirs.) This also agrees with solar-type field binaries \citep{Raghavan2010,Moe2017}, except that field binaries also show an excess of ``twins'' (with $q=1$).  Given the precision on our photometric $q$ measurements, we are unable to probe our data for a twin excess.    

\section{Discussion}\label{sec:discuss}

In this paper, we self-consistently investigate the photometric binaries in six well-studied OCs (NGC 2168, NGC 7789, NGC 6891, NGC 2682, NGC 188, NGC 6791) as a precursor to a larger endeavor including many (less studied) OCs.  The primary goals of this project are to ($i$) compare the binary populations  across different OCs (e.g., to see if different primordial binary populations are required by the data in different OCs), ($ii$) compare the binaries in OCs  with the field (because many OCs are believed to dissolve quickly to populate the field), and ($iii$) investigate empirically how the stellar dynamical environment impacts the binary populations in OCs (e.g., to guide detailed $N$-body models).  In this section, we discuss each of these points in turn for the sample of clusters studied here.

\subsection{Consistency in binary populations across different OCs} 

We find that the overall photometric binary fractions in the OCs studied here and within our ML-MS sample ($0.6 M_\odot < M_1 < 1 M_\odot$) range from about 30\% to 60\% (Table~\ref{tab:binary_results}).  NGC 2168, NGC 6819, NGC 2682 and NGC 188 have statistically indistinguishable binary fractions, while NGC 7789 is significantly lower and NGC 6791 is significantly higher.  Both of these clusters are somewhat difficult to analyze. NGC 6791 is distant and dense, and therefore the stars in our sample are relatively faint and have large photometric uncertainties. Possibly the uncertainty on the binary fraction resulting from Poisson counting statistics (as included in Table~\ref{tab:binary_results}) underestimates the true uncertainty on the binary fraction derived from our analysis, especially for NGC 6791.  This may result in additional stars being considered binaries that perhaps should be singles or field stars.  
NGC 7789 is embedded in a very rich field (making it challenging to separate cluster from field stars), has relatively high reddening (which we were unable to correct for in Section~\ref{ssec:red}), and it appears that the isochrone does not fit the lower main sequence in this cluster particularly well (Figure~\ref{fig:CMDs}).  This last point may result in a lower binary fraction, as it appears that the isochrone moves too far to the red (thereby underestimating the binary mass ratios and likely considering some real binaries as single stars).  Nonetheless, the binary fractions in our full OC sample follow a suggestive trend with cluster age (which persists even if we exclude NGC 7789 and NGC 6791, see Section~\ref{ssec:bexternal} and below). 

The similarity in the photometric binary fractions for NGC 2168, NGC 6819, NGC 2682 and NGC 188 is also in agreement with results from the WOCS spectroscopic surveys.  Though the WOCS binary fractions are self-similar, our photometric binary fractions for these clusters are consistently higher than those in the WOCS survey, likely due to long-period binaries detected here that are beyond the reach of WOCS time-series RV measurements.   

We find that the binary fractions in these OCs are independent of metallicity (Figure~\ref{fig:fb}).  \cite{Cordoni2023} also did not find a significant correlation between binary fraction and OC metallicity when considering the binary fractions in 78 OCs.  This lack of correlation is also seen in GCs \citep{Milone2012, Ji201}. 


Though we find that BASE-9 does not provide high-precision $q$ values with these photometric data sets and OCs, it appears that all of the OCs studied here have $q$ distributions consistent with being uniform.  This result is also consistent with findings from previous WOCS spectroscopic studies of M67 and NGC 188 (while the other clusters do not have similar analyses in the literature) and with findings from \cite{Milone2012} who found that the mass ratio distribution is flat for $q>0.5$.

\subsection{Comparison of OC and field binaries}

Broadly, the OC binaries studied here are consistent with similar binaries in the field.  In the upper-left panel of Figure~\ref{fig:fb}, we compare the solar-type field binary ($\sim$30\%) and multiplicity ($\sim$50\%) fractions \citep{Raghavan2010} with those in our OCs.  We include both the binary and multiplicity fractions because some of our photometric ``binaries'' may in fact be higher-order multiple systems that have low enough mass ratios to still reside in the binary locus on a CMD.  All our clusters are consistent with this range in binary/multiplicity fraction of the field, to within the uncertainties.  

The field multiplicity fraction is also observed to have a significant correlation with primary star mass \citep{Raghavan2010, Duchene2013}.  We investigate this trend in the left panel of Figure~\ref{fig:all_fbs}.  Though visually there appears to be a hint of a similar trend as seen in the field, formally we find a slope of binary fraction vs.\ primary mass of $0.14 \pm 0.11$, consistent with zero (though intriguingly similar in value to the slope found for a similar sample of binaries in GCs by \citealt{Milone2012}). Within our mass range, the field binaries in the literature essentially only have one mass bin; the trend with mass in the field comes from binaries with mass significantly smaller and larger than those studied here.  A larger OC sample, with a broader primary mass range, will be required to investigate this further. 

The $q$ distribution of the OCs studied here are also broadly consistent with solar-type field stars, which are observed to have a roughly uniform distribution.  The field binaries also show evidence for an increased frequency of ``twins'' relative to lower-$q$ systems, which we are currently unable to probe with our data. 

\subsection{Impacts of stellar dynamics and relaxation}

Star clusters are expected to become mass segregated over the course of a few relaxation timescales.  Most of the OCs in our sample have survived through at least one relaxation time (Section~\ref{ssec:trelax} and Table~\ref{tab:systems}), and indeed these OCs show evidence for mass segregation (though some show stronger evidence than others, see Figures \ref{fig:M1_CDFs}  and \ref{fig:MLMS_CDFs} and Section~\ref{ssec:binternal}).  Interestingly, though NGC 2168 has not yet reached a full relaxation time, it does show evidence of mass segregation, particularly for the binaries (as is studied in more detail in \cite{Motherway2023}).  The effects of mass segregation are also evident in the right panel of Figure~\ref{fig:all_fbs}, where the binary fraction increases dramatically toward the cluster centers in our sample.  This agrees with findings for the WOCS RV binaries in our sample \citep[e.g.,][]{Geller2012, Geller2021} and also for similar binaries in GCs \citep{Milone2012}.

Mass segregation and cluster evaporation may also lead to the global binary fraction increasing with time, as lower-mass single stars are preferentially lost from the cluster.  In our data, we observe a significant trend of increasing binary fraction with increasing cluster age (see upper-left panel of Figure~\ref{fig:fb} and Section~\ref{ssec:bexternal}).  However, we do not find any trend of binary fraction with number of relaxation times. Some previous studies of photometric binaries in OCs have also observed such a trend between binary fraction and cluster age \citep{Donada2023}, while others have not \citep{Pang2023} (see also Section \ref{sec:OC_v_GCs}).  However, many of these previous studies employ less reliable techniques to identify photometric binaries (i.e., ``chi-by-eye"). Furthermore, $N$-body models suggest that the overall solar-type binary fraction in a cluster may remain relatively constant until the very end of cluster's life when it begins to dissolve more rapidly \citep{gellernbody}.  We will investigate this trend with a larger data set including additional OCs in a future paper.

Close gravitational encounters can also significantly modify binaries, e.g., through destruction of soft binaries and also exchanges.  Binary destruction is expected to be most prevalent within denser clusters and those with higher velocity dispersions, where encounters are more frequent and more energetic.  In the upper-right panel of Figure~\ref{fig:fb} we plot the binary fraction against cluster central density, and in the lower-right panel of Figure~\ref{fig:fb} we plot the binary fraction against cluster mass (which can be a proxy for velocity dispersion).  Including all OCs in our sample, we find no significant trend with either parameter.  However, we note that NGC 6791 is the densest, most metal rich, oldest and tied for the most massive cluster in our sample, and therefore may have undue weight in these correlations.  If we remove NGC 6791 we see that the binary fraction is significantly anti-correlated with central density and cluster mass.  Note that these correlations also rely on NGC 7789, which may have an underestimated binary fraction.  Therefore, although these correlations are suggestive, and may be the result of increased stellar encounters resulting in the destruction of binaries (or perhaps increased initial cluster density prohibiting the formation of wide binaries), more data are needed before drawing firm conclusions. 

\subsection{Comparison of OC and GC binaries} \label{sec:OC_v_GCs}
Due to the higher density of GCs, dynamics are expected to affect the binary populations more severely, e.g.\ by a larger frequency of close stellar encounters and higher-energy in such encounters on average.  Such pronounced dynamical evolution can result in the increased destruction of wide binaries and increased exchanges (leading toward equal mass ratios), among other effects.

While we find a significant positive correlation between OC age and binary fraction, this is the opposite of what is observed in GCs, which show evidence of a decrease in binary fraction with cluster age \citep{Sollima2007,Milone2012, Ji201}.  On the other hand, neither OCs or GCs show a correlation between binary fraction and dynamical age, which one might expect if dynamics are responsible.
\citet{Ji201} suggests this may be the result of different primordial binary populations (where younger GCs are born with higher binary fractions). We will investigate this discrepancy between OCs and GCs further in the future when we expand our study to a larger dataset of OCs. 

\cite{Ji201} and \cite{Milone2012} also found a negative, though weak, correlation between GC luminosity (mass) and binary fraction, which may also be tied to dynamics. In our OC data, we only see a trend in binary fraction with mass when we exclude NGC 6791.  

OCs and GCs also both show no correlation with cluster metallicity \citep{Milone2012, Ji201}, and the $q$ distribution for $q>0.5$ is consistent with uniform in both environments.  Future work to expand our OC data set will enable further comparisons to the more dynamically active environments of GCs.

\section{Conclusions}\label{sec:Conclusions}
We present results from a self-consistent Bayesian analysis of the photometric data for six OCs (NGC 2168, NGC 7789, NGC 6819, NGC 188 and NGC 6791) that span a wide range of ages, masses, and metallicity.  We use Gaia kinematics and distances to select a sample of likely cluster members.  We then utilize the Bayesian software suite BASE-9 with photometric data, in eleven filters from the Gaia, Pan-STARRS, and 2MASS surveys, along with PARSEC stellar evolution models to derive posterior distributions of global cluster parameters (age, distance, metallicity and reddening), individual stellar masses, binary mass ratios, and photometric membership probabilities.  From these results, we identify a sample of cluster members, including a rich population of (photometric) binary stars in each cluster.

We perform a careful completeness analysis using simulated clusters created with BASE-9 and analyze them with the same pipeline as our real data.  We find that our recovery rate for binaries with $q > 0.5$ is nearly complete, and that below this mass ratio our completeness drops off considerably.   
We also compare our results to those from the WOCS spectroscopic surveys of these clusters and find that BASE-9 identifies many of the WOCS binaries that we find to be Gaia members (and does best in clusters with low reddening). Many that are missed lie far from standard isochrones, which BASE-9 is unable to model, or likely have low mass ratios that BASE-9 is not sensitive to.  We are therefore confident that our technique is robust.

We find that for most of the OCs in our sample, the binary fractions are consistent, at roughly 40-50\%, which is also broadly in agreement with the field multiplicity fraction.  Also similar to the field, we find that the distributions in $q$ for these OCs are consistent with being uniform.  We also find a hint of a correlation between binary fraction and primary-star mass (as also seen in the field), but data from more OCs are required for verification.

Within the OCs, the binary fraction increases dramatically toward the cluster centers, likely due to mass segregation.  The effects of mass segregation are also evident within the single and binary samples themselves in most of the OCs studied here, where higher-mass populations tend to be more centrally concentrated than lower-mass populations.    
Interestingly, we find that the binary fraction in our sample of OCs is also significantly correlated with cluster age (though not age/$t_{rh}$), such that the older clusters in our sample have higher binary fractions. This could potentially also be related to mass segregation and it's interplay with cluster dissolution, where lower-mass single stars are preferentially lost from the clusters over time.  

We also observe a hint of an anti-correlation between binary fraction and cluster central density and cluster mass, respectively, though more data in additional OCs are needed to verify these trends.   If this result persists, it may be evidence for dynamical interactions destroying binaries (or prohibiting their formation), as has long been predicted in numerical models.


With these six well-studied OCs we have developed a standardized and self-consistent pipeline to derive binary characteristics from photometric, kinematic and astrometric data in OCs.  In future papers, we aim to use this technique to study the binary populations in a much larger sample of (less studied) OCs to increase our understanding of the relationship between a binary population and its host cluster characteristics and to test theoretical predictions from star cluster models.

\begin{acknowledgments}
We thank Elliot Robinson, Elizabeth Jeffery, Emily Leiner, David Stenning, David van Dyk, and Bill Jefferys for helpful conversations that have improved this manuscript.  ACC and AMG acknowledge supported from the National Science Foundation (NSF) under grant No. AST-2107738. Any opinions, findings, and conclusions or recommendations expressed in this material are those of the author(s) and do not necessarily reflect the views of the NSF. This research was supported in part through the computational resources and staff contributions provided for the Quest high performance computing facility at Northwestern University which is jointly supported by the Office of the Provost, the Office for Research, and Northwestern University Information Technology.
\end{acknowledgments}

%

\vspace{5mm}
\facilities{Gaia DR3 (DOI:10.5270). The Pan-STARRS1 Surveys (PS1) and the PS1 public science archive have been made possible through contributions by the Institute for Astronomy, the University of Hawaii, the Pan-STARRS Project Office, the Max-Planck Society and its participating institutes, the Max Planck Institute for Astronomy, Heidelberg and the Max Planck Institute for Extraterrestrial Physics, Garching, The Johns Hopkins University, Durham University, the University of Edinburgh, the Queen's University Belfast, the Harvard-Smithsonian Center for Astrophysics, the Las Cumbres Observatory Global Telescope Network Incorporated, the National Central University of Taiwan, the Space Telescope Science Institute, the National Aeronautics and Space Administration under Grant No. NNX08AR22G issued through the Planetary Science Division of the NASA Science Mission Directorate, the National Science Foundation Grant No. AST-1238877, the University of Maryland, Eotvos Lorand University (ELTE), the Los Alamos National Laboratory, and the Gordon and Betty Moore Foundation.  This publication makes use of data products from the Two Micron All Sky Survey, which is a joint project of the University of Massachusetts and the Infrared Processing and Analysis Center/California Institute of Technology, funded by the National Aeronautics and Space Administration and the National Science Foundation.}


\software{astropy \citep{2013A&A...558A..33A,2018AJ....156..123A},  
dustmaps: A Python interface for maps of interstellar dust \citep{Green2018}          }


\bibliography{main}{}
\bibliographystyle{aasjournal}

\appendix

\section{Additional CMDs}\label{sec:CMDs}
We show CMDs for the six OCs using 2MASS (Figure \ref{fig:2M_CMDs}) and Pan-STARRS photometry (Figure \ref{fig:PS_CMDs} and \ref{fig:PS_zy_CMDs}) below.  In Figure \ref{fig:synth_CMD} we show synthetic CMDs using Gaia (left), Pan-STARRS (middle), and 2MASS (right) filters.  This synthetic photometry is generated using the isochrone for NGC 2168 which is shown in red.  We generate only data along the magnitudes that correspond to $0.6 \, M_{\odot} \leq M_{1} \leq 1 \, M_{\odot}$.  These synthetic data are used to test for incompleteness, as described in Section \ref{sec:incompleteness}, and are generated in each filter of Gaia, Pan-STARRS and 2MASS for each of the six OCs.  The BASE-9 determined single stars are shown in dark gray and the binary stars are colored according to their $q_{\rm in}/q_{\rm out}$ value.  $q_{\rm in}$ is the true, simulated $q$ value and $q_{\rm out}$ is the $q$ value as determined by BASE-9.
\begin{figure*}[!ht]
	\includegraphics[width=\columnwidth]{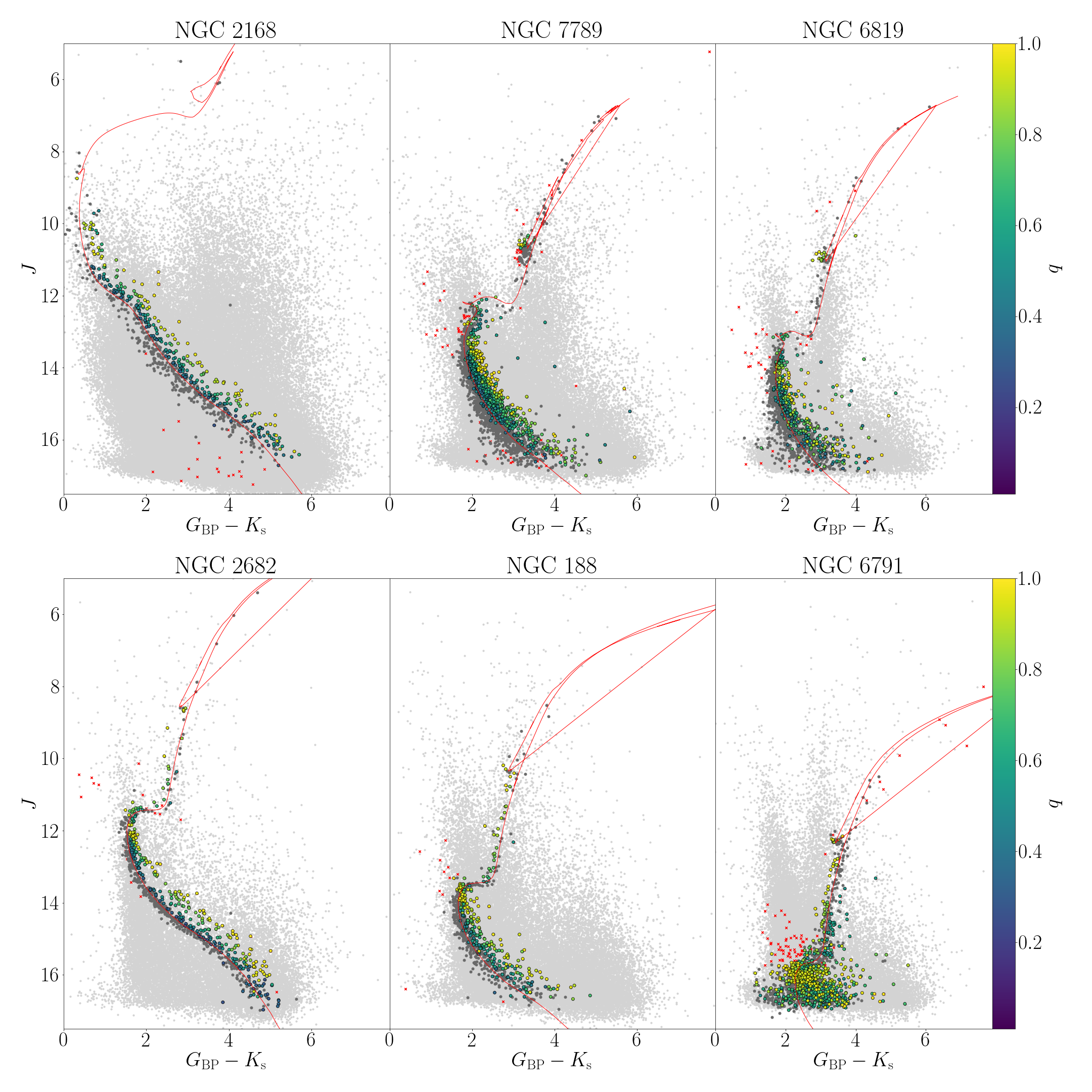}
    \caption{CMDs with 2MASS $JK_{\rm s}$ photometry showing BASE-9 results for each cluster.  The symbols are the same as denoted in Figure \ref{fig:CMDs}.  Field stars are marked in light gray and stars classified as members using Gaia kinematics and distances but are rejected by BASE-9 are marked by red 'x's.  Single star BASE-9 members are marked in dark gray and BASE-9 binary members are colored according to the mass ratio of the binary.  The red line shows a PARSEC isochrone created from the median cluster parameters from our BASE-9 analysis.} 
    \label{fig:2M_CMDs}
\end{figure*}

\begin{figure*}[!ht]
	\includegraphics[width=\columnwidth]{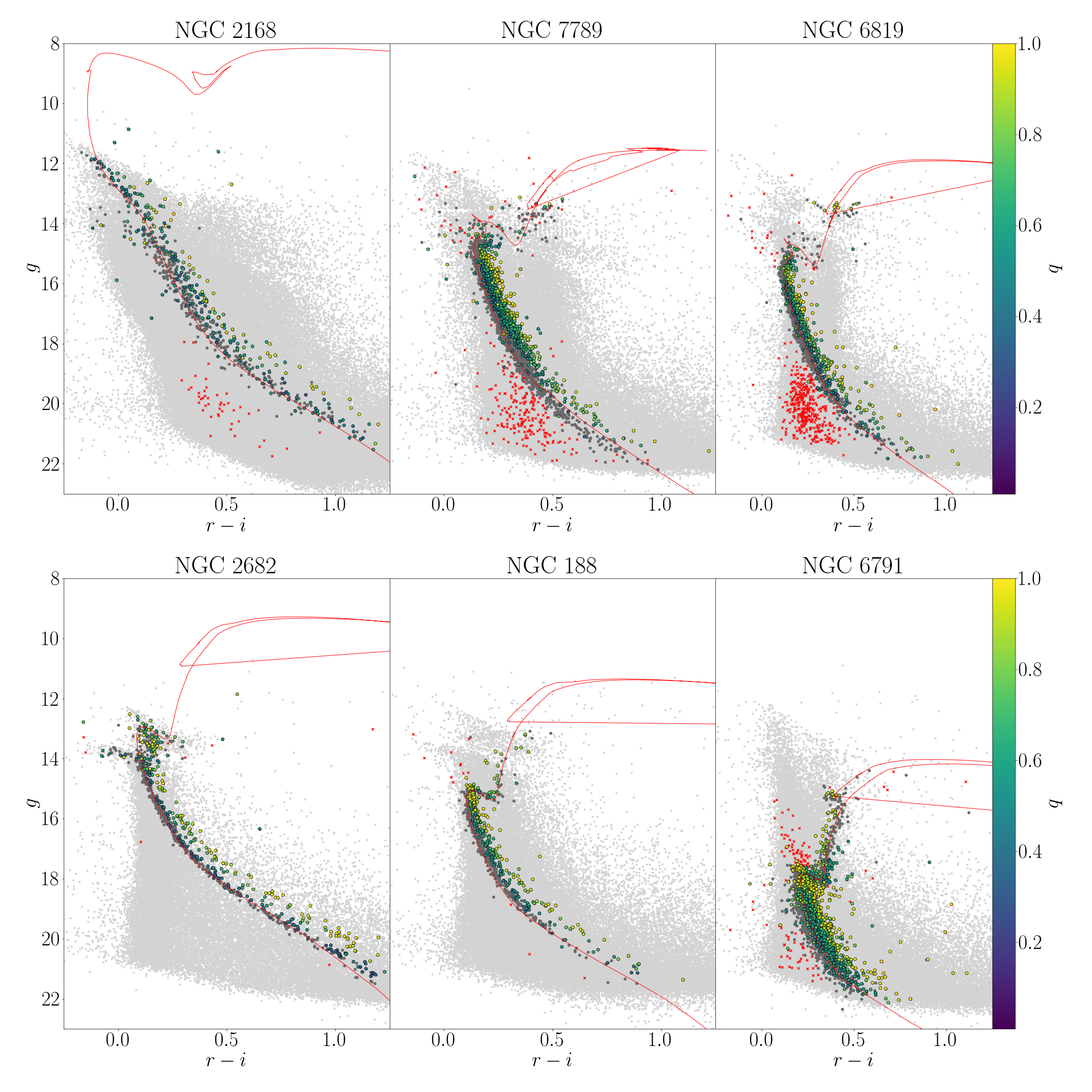}
    \caption{CMDs with Pan-STARRS $g,r,$ and $i$ photometry showing BASE-9 results for each cluster.  The symbols are the same as denoted in Figure \ref{fig:CMDs}.  Field stars are marked in light gray and stars classified as members using Gaia kinematics and distances but are rejected by BASE-9 are marked by red 'x's.  Single star BASE-9 members are marked in dark gray and BASE-9 binary members are colored according to the mass ratio of the binary.  The red line shows a PARSEC isochrone created from the median cluster parameters from our BASE-9 analysis.} 
    \label{fig:PS_CMDs}
\end{figure*}

\begin{figure*}[!ht]
	\includegraphics[width=\columnwidth]{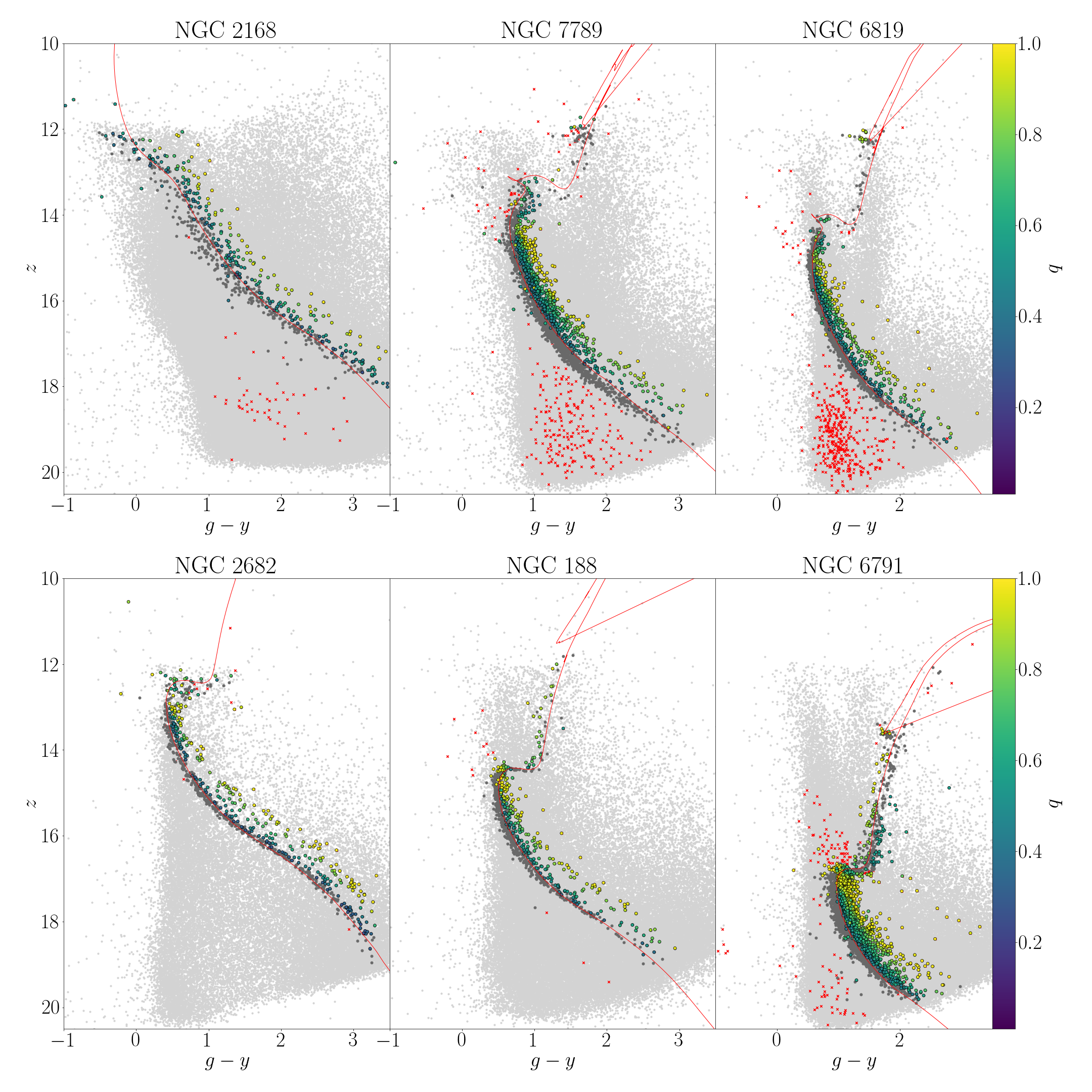}
    \caption{CMDs with Pan-STARRS $g,z,$ and $y$ photometry showing BASE-9 results for each cluster.  The symbols are the same as denoted in Figure \ref{fig:CMDs}.  Field stars are marked in light gray and stars classified as members using Gaia kinematics and distances but are rejected by BASE-9 are marked by red 'x's.  Single star BASE-9 members are marked in dark gray and BASE-9 binary members are colored according to the mass ratio of the binary.  The red line shows a PARSEC isochrone created from the median cluster parameters from our BASE-9 analysis.} 
    \label{fig:PS_zy_CMDs}
\end{figure*}

\begin{figure*}[!ht]
	\includegraphics[width=\columnwidth]{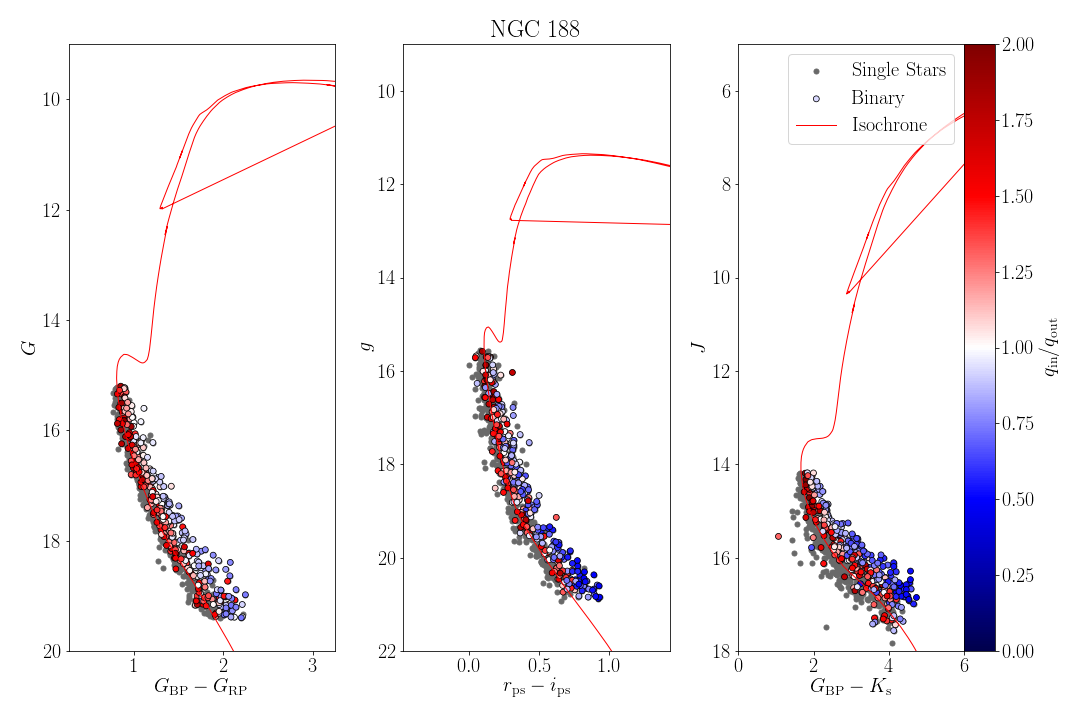}
    \caption{Synthetic CMDs with Gaia (left), Pan-STARRS (middle), and 2MASS $J$ and $K_{\rm s}$ filters (right).  This synthetic photometry is generated using the isochrone for NGC 188 which is shown in red.  The BASE-9 determined single stars are shown in dark gray and the binary stars are colored according to their $q_{\rm in}/q_{\rm out}$ value.} 
    \label{fig:synth_CMD}
\end{figure*}

\section{Differential reddening}\label{sec:diff_red}
\begin{figure*}[!ht]
	\includegraphics[width=\columnwidth]{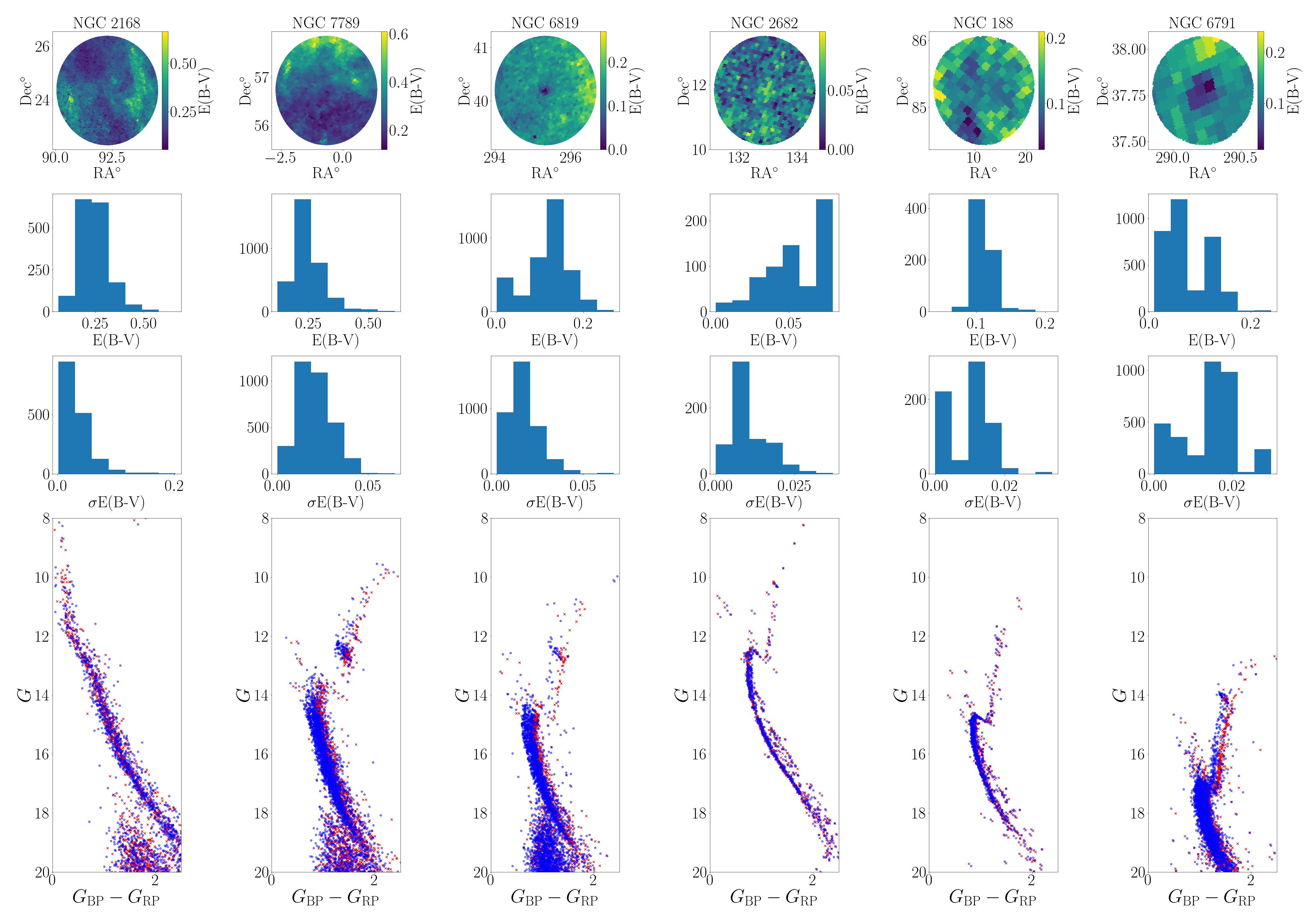}
    \caption{$E(B-V)$ \texttt{Bayestar19} reddening map for the area of sky in which each cluster resides (top row), histograms of the $E(B-V)$ values for the likely cluster members as determined from Gaia kinematics and distances (second from top row), histograms of the uncertainties on these $E(B-V)$  values (third from top row), and CMDs in Gaia filters for the original photometry in red and the photometry after correcting for differential reddening in blue (bottom row).} 
    \label{fig:diff_redd}
\end{figure*}

In Figure \ref{fig:diff_redd} we show (top row) the $E(B-V)$ \texttt{Bayestar19} reddening map for the area of sky in which each cluster resides, (second row from top) histograms of the $E(B-V)$ values for the likely cluster members as determined from Gaia kinematics and distances, and (third row from top) histograms of the uncertainties on these values.  For each cluster, this same set of stars is shown in a CMD in Gaia filters for the original photometry in red and the photometry after correcting for differential reddening in blue, in the bottom row.  In our analysis, we use the photometric data sets that have been corrected for differential reddening for all clusters except NGC 7789 and NGC 6819.  We found BASE-9 had difficulty sampling and deriving reliable cluster parameters for these clusters when using the corrected photometric data sets.  Aside from NGC 2168, these two clusters have the largest reddening values and associated reddening errors.  We suspect that BASE-9's inability to recover reliable cluster parameters using the corrected photometric data sets for these clusters is likely due to the relatively large photometric errors introduced by these reddening corrections.


\end{document}